\pdfoutput=1
\documentclass[a4paper,10pt]{article}
\usepackage{jheppub,amsmath,amsfonts,amssymb,xcolor,graphicx,hyperref}
\usepackage[utf8x]{inputenc}
\usepackage[section]{placeins}
\hypersetup{colorlinks=true,linkcolor=blue,citecolor=magenta,filecolor=magenta,
  urlcolor=cyan}
\def\ie{{\it i.e.}}
\newcommand{\ch}{\textsc{CalcHep}}
\newcommand{\del}{\textsc{Delphes}~3}
\newcommand{\fj}{\textsc{FastJet}}
\newcommand{\fr}{\textsc{FeynRules}}
\newcommand{\hb}{\textsc{HiggsBounds}}
\newcommand{\hs}{\textsc{HiggsSignals}}
\newcommand{\mdm}{\textsc{MadDM}}
\newcommand{\mg}{\textsc{MG5\_aMC}}
\newcommand{\ma}{\textsc{MadAnalysis}~5}
\newcommand{\mo}{\textsc{MicrOMEGAs}}
\newcommand{\py}{\textsc{Pythia}~8}
\newcommand{\pysl}{\textsc{PySLHA}}
\newcommand{\be}{\begin{equation}}
\newcommand{\ee}{\end{equation}}
\def\bsp#1\esp{\begin{split}#1\end{split}}
\def\bpm{\begin{pmatrix}}
\def\epm{\end{pmatrix}}
\newcommand{\bea}{\begin{eqnarray}}
\newcommand{\eea}{\end{eqnarray}}

\def\gL{g_{\scriptscriptstyle L}}
\def\gR{g_{\scriptscriptstyle R}}
\def\gBL{g_{\scriptscriptstyle B-L}}
\def\gY{g_{\scriptscriptstyle Y}}
\def\gF{G_{\scriptscriptstyle F}}
\def\vR{v_{\scriptscriptstyle R}}
\def\vL{v_{\scriptscriptstyle L}}
\def\tw{\theta_{\scriptscriptstyle W}}
\def\phw{\varphi_{\scriptscriptstyle W}}
\def\zw{\vartheta_{\scriptscriptstyle W}}
\def\lag{{\cal L}}

\title{Natural dark matter and light bosons with an alternative left-right
  symmetry}

\author[a]{Mariana~Frank}
\author[b,c]{\!\!, Benjamin~Fuks}
\author[a]{\! and \"{O}zer \"{O}zdal}

\emailAdd{mariana.frank@concordia.ca}
\emailAdd{fuks@lpthe.jussieu.fr}
\emailAdd{ozer.ozdal@concordia.ca}

\affiliation[a]{Department of Physics, Concordia University,
  7141 Sherbrooke St. West, Montreal, Quebec, Canada H4B 1R6}
\affiliation[b]{Laboratoire de Physique Th\'eorique et Hautes Energies (LPTHE),
  UMR 7589, Sorbonne Universit\'e et CNRS, 4 place Jussieu,
  75252 Paris Cedex 05, France}
\affiliation[c]{Institut Universitaire de France, 103 boulevard Saint-Michel,
   75005 Paris, France}

\abstract{
We perform a consistent analysis of the 
alternative left-right
symmetric model emerging from $E_6$ grand unification. We include a large set of
theoretical and experimental constraints, with a particular emphasis on dark
matter observables and collider signals. We show that the exotic neutrino
inherent to this class of models, the scotino, is a viable candidate for dark
matter satisfying relic density and direct detection constraints. This has
strong implications on the scotino mass restricting it to lie in a narrow
window, as well as on the spectrum of Higgs bosons,  rendering it predictable,
with a few light scalar, pseudoscalar and charged states. Moreover, we also show
that the extra charged $W'$ gauge boson can be light, and investigate the most
promising signals 
at the future high-luminosity upgrade of the
LHC. Our findings show that the most optimistic cosmologically-favoured scenarios should be observable at $5\sigma$, whilst others could leave visible hints provided the background is under good control at the systematical level.
}

\begin{document}
\preprint{CUMQ/HEP 202}

\maketitle
\flushbottom

\section{Introduction}
\label{sec:intro}
The nature of dark matter and its interactions 
is one of the most
puzzling conceptual issues of the Standard Model of particle physics and points
clearly towards the existence of new physics. So far, the most popular
extensions of the Standard Model (SM) that contain natural dark matter (DM)
candidates have been either supersymmetric, so that $R$-parity conservation
enforces a stable supersymmetric state behaving as a weakly-interacting massive
particle (WIMP)~\cite{Arcadi:2017kky}, or featuring axion-like particles that
could additionally shed light on a potential solution to the strong $CP$
problem~\cite{Kawasaki:2013ae,Graham:2015ouw}. While experimental DM searches
are on-going and put stronger and stronger constraints on the phenomenological
viability of the models, several new {\it ad-hoc} mechanisms have been recently
designed  to supplement the SM with a DM candidate. In the latter, the observed properties of
DM~\cite{Aghanim:2018eyx} can be successfully reproduced by an appropriate
tuning of the particle masses and properties. For instance, new force carriers
could be introduced to mediate the interactions of the dark sector with the SM
one, as within the dark photon or vector portal models~\cite{Feldman:2006wd,
Izaguirre:2015yja,Curtin:2014cca,Essig:2013lka,Davoudiasl:2012ag}. Differently,
the connection between the dark and visible sector could be realised through
interactions with vector-like fermions~\cite{Toma:2013bka,Giacchino:2013bta,
Giacchino:2014moa,Ibarra:2014qma,Giacchino:2015hvk,Colucci:2018vxz}. Whilst
appealing from a phenomenological point of view by virtue of their simplicity,
such DM setups are however quite unnatural. In this work, we therefore go back
to natural dark matter models and focus on a less studied class of scenarios
that emerges from the grand unification of the SM gauge interactions.

Grand unification models based on the breaking of the exceptional group
$E_6$~\cite{Gursey:1975ki,Achiman:1978vg} have been popular for awhile, at the
beginning as a result of developments in string theories~\cite{Hewett:1988xc},
then later as generators of models with additional $U(1)$ symmetries~\cite{
Langacker:1998tc}. These so-called $U(1)'$ models arise from considering the
$SO(10)\times U(1)$ subgroup of $E_6$. However, the $E_6$ group has also an
$SU(3)\times SU(3)\times SU(3)$ subgroup. One of these $SU(3)$ remains unbroken
and is associated with the SM strong interaction group $SU(3)_c$, while the
two others further break into the $SU(2)_L \times SU(2)_H \times U(1)_X$ group
that embeds the $SU(2)_L \times U(1)_Y$ electroweak symmetry. In the so-called
left-right symmetric model (LRSM), that naturally accounts for non-vanishing
neutrino masses~\cite{Pati:1974yy,Mohapatra:1974gc,Senjanovic:1975rk,
Mohapatra:1977mj}, $SU(2)_H$ is identified with $SU(2)_R$ and
$U(1)_X$ with $U(1)_{B-L}$. In such a configuration, the right-handed SM
fermions and the right-handed neutrino $\nu_R$ are collected into $SU(2)_R$
doublets. The structure of the Higgs sector could however lead to non-acceptable
tree-level flavour-violating interactions that would conflict with the observed
properties of kaon and $B$-meson systems. Consequently, the $SU(2)_R \times
U(1)_{B-L}$ symmetry has to be broken at a very high energy scale to
mass-suppress any potential flavour-violating effect. This additionally pushes
the masses of the extra Higgs and gauge bosons of the model to the high scale,
making them unlikely to detect at the LHC. Furthermore, in its minimal
incarnation, the LRSM lacks any viable DM candidate~\cite{Bahrami:2016has}.

It is nevertheless possible to associate the $SU(2)_H$ symmetry with a different
$SU(2)_{R'}$ group in which the assignments of the SM fermions into doublets
are different~\cite{Ma:1986we,Frank:2005rb}. This model is called the
alternative left-right symmetric model (ALRSM)~\cite{Babu:1987kp,Ma:2010us}. In
this case, the $SU(2)_{R'}$ partner of the right-handed up-quark $u_R$ is an
exotic down-type quark $d_R'$ (instead of the SM right-handed down-type quark
$d_R$), and the $SU(2)_{R'}$ partner of the right-handed charged lepton $e_R$ is
a new neutral lepton, the scotino $n_R$ (instead of the more standard
right-handed neutrino $\nu_R$). The right-handed neutrino $\nu_R$ and down-type
quark $d_R$ therefore remain singlets under both the $SU(2)_L$ and $SU(2)_{R'}$
groups. In addition, the model field content also includes $SU(2)_L$ singlet
counterparts to the new states, \ie\ an $n_L$ scotino and a $d_L'$ down-type
quark. Consequently, one generation of quarks is described by one $SU(2)_L$
doublet $Q_L = (u_L, d_L)$, one $SU(2)_{R'}$ doublet $Q_R = (u_R, d_R')$ and two
$SU(2)_L\times SU(2)_{R'}$ singlets $d_L'$ and $d_R$. Similarly, one generation
of leptons is described by one $SU(2)_L$ doublet $L_L = (\nu_L, e_L)$, one
$SU(2)_{R'}$ doublet $L_R = (n_R, e_R)$ and two $SU(2)_L\times SU(2)_{R'}$
singlets $n_L$ and $\nu_R$. Moreover, the right-handed neutrino $\nu_R$ and the $n_L$
scotino being singlets under $U(1)_{B-L}$,  are unlikely to be
viable DM candidates, as their too weak interactions with the SM particles would
make them over-abundant. On the contrary, the $n_R$ scotino may fulfill the role.

In this work, we will show that this is indeed the case. The $n_R$ scotino can
be an acceptable DM candidate satisfying requirements from imposing agreement
with the observed relic density and the non-violation of the DM direct and indirect detection
bounds. This however yields very stringent constraints on the model parameter
space. In contrast with the usual LRSM, the charged right-handed gauge boson
$W'$ couples right-handed up-type quarks and charged leptons to their exotic
quarks and scotino partners. Therefore, the limits on the $W'$-boson mass
(originating mainly from the properties of the $K^0-{\bar K}^0$ mixing in the
LRSM case~\cite{Barenboim:1996nd}) do not apply. Similarly, the different
couplings of the Higgs states to fermions  forbid most dangerous
flavour-violating effects, so that the mass limits on the Higgs states can also
be relaxed. As will be demonstrated in the rest of this paper, these
considerations lead to a quite predictable lower-energy spectrum with signatures
potentially observable at the high-luminosity LHC.

The aim of this work is therefore to provide a comprehensive analysis of the
ALRSM setup, emphasising for the first time the complementarity between
cosmological, low-energy and collider constraints in this class of extensions of
the SM. We update and extent previous recent works that have focused on the dark
matter~\cite{Khalil:2010yt} and collider~\cite{Ashry:2013loa} phenomenology
independently. In section~\ref{sec:ALRSMmodel}, we provide a brief description
of the ALRSM and detail the technical setup underlying our analysis in
 section \ref{sec:scan}. Our results are presented in the next sections. In
section~\ref{sec:gaugebosons}, we analyse the constraints on the model parameter
space originating from LHC searches for new gauge bosons,  performed in a
similar way as for the LRSM~\cite{Frank:2018ifw}. Section~\ref{sec:dm}
is dedicated to cosmological considerations and their impact on the parameter
space. In section~\ref{sec:collider} we focus on determining promising
signals of the model at the future high-luminosity upgrade of the LHC. We
summarise our work and conclude in section~\ref{sec:conclusion}. In
appendices~\ref{app:higgs} and \ref{app:ferms}, we include further details on
the diagonalisation of the model Higgs and fermionic sector respectively, and
document our implementation of the ALRSM in \fr~\cite{Alloul:2013bka} in
appendix~\ref{app:fr}.

\section{The alternative left-right symmetric model}
\label{sec:ALRSMmodel}
The alternative left-right symmetric model~\cite{Ma:1986we,Babu:1987kp,
Frank:2005rb,Ma:2010us} is a variant of the more usual minimal left-right
symmetric model. It is based on the $SU(3)_c \times SU(2)_L \times
SU(2)_{R'} \times U(1)_{B-L}$ gauge group, to which we supplement a global
$U(1)_S$ symmetry. The spontaneous breaking of $SU(2)_{R'}\times U(1)_S$ is
implemented so that the $L=S+T_{3R}$ charge, that can be seen as a generalised
lepton number, remains unbroken (with $T_{3R}$ being the third generator of
$SU(2)_{R'}$).

The quantum numbers and representations chosen for the fermionic field content
of the ALRSM are motivated by heterotic superstring models in which all SM
matter multiplets are collected into a ${\bf 27}$-plet of $E_6$.
Under the $E_6$ maximal subgroup $SU(3)_c \times SU(3)_L \times SU(3)_H$,
the ${\bf 27}$ representation is decomposed as
\be
  {\bf  27} =  \big({\bf 3}, {\bf 3}, 1\big) +
    \big({\bf \bar 3}, 1, {\bf \bar 3}\big) +
    \big(1, {\bf \bar 3}, {\bf 3}\big)
   \qquad \equiv \quad q \quad + \quad {\bar q} \quad + \quad l \ .
\ee
Explicitly, the particle content for this decomposition can be written, ignoring
the sign structure for clarity, as
\be\renewcommand{\arraystretch}{1.4}
  q        = \bpm u_L\\d_L\\d_L'\epm \ , \qquad
  {\bar q} = \bpm u_R^c & d^c_R & d_R^{\prime c}\epm \ , \qquad
  l = \bpm E_R^c&N_L&\nu_L\\ N_R^c&E_L&e_L\\ e_R^c&\nu_R^c&n_R^c\epm\ ,
\label{eq:331_fields}\ee
where $d'$, $E$, $N$ and $n$ are exotic fermions and $u$, $d$, $e$ and $\nu$ are
the usual up-type quarks, down-type quarks, charged leptons and neutrinos. In
this setup, $SU(3)_L$ operates vertically and $SU(3)_H$ horizontally. There are
three different ways to embed $SU(2)_H$ into $SU(3)_H$~\cite{Ma:1986we}. The
most common one consists in imposing the first and second column of the above
multiplets to form $SU(2)_H$ doublets, which corresponds to the usual LRSM
($SU(2)_H = SU(2)_R$)~\cite{Pati:1974yy,Mohapatra:1974gc,Senjanovic:1975rk,
Mohapatra:1977mj}. The second option requires in contrast that the first
and third columns of the above multiplets form an $SU(2)_H$ doublet, which
corresponds to the ALRSM ($SU(2)_H = SU(2)_{R'}$)~\cite{Ma:1986we,Babu:1987kp,
Frank:2005rb,Ma:2010us} . Finally, the third and last option corresponds to
doublets formed from the second and third columns of the above multiplets, which
corresponds to the Inert Doublet Model ($SU(2)_H = SU(2)_I$)~\cite{
Deshpande:1977rw,Ma:2006km,Barbieri:2006dq}.

We are interested here in the
second option. In the rest of this section, we present a summary of the model
description, leaving computational details for the appendix. While previous
descriptions of the ALRSM exist, we provide extensive details to properly and
consistently define our notations, which is relevant for the model
implementation in the high-energy physics tools depicted in
section~\ref{sec:scan}.

\begin{table}
  \renewcommand{\arraystretch}{1.2}\setlength\tabcolsep{5pt}
  \begin{center}
    \begin{tabular}{ccc}
      Fields & Repr.&$U(1)_S$\\\hline\hline\\[-.4cm]
      $Q_L = \bpm u_L\\d_L\epm$ &
        $\big({\bf 3}, {\bf 2}, {\bf 1},  \frac16\big)$ & 0 \\[.4cm]
      $Q_R = \bpm u_R\\d_R'\epm$ &
        $\big({\bf 3}, {\bf 1}, {\bf 2},  \frac16\big)$ & $-\frac12$\\[.4cm]
      $d'_L$ & $\big({\bf 3}, {\bf 1}, {\bf 1}, -\frac13\big)$ & $-1$\\[.2cm]
      $d_R$  & $\big({\bf 3}, {\bf 1}, {\bf 1}, -\frac13\big)$ & 0\\[.2cm]
    \hline\\[-.4cm]
      $L_L = \bpm \nu_L\\e_L\epm$ &
        $\big({\bf 1}, {\bf 2}, {\bf 1},  -\frac12\big)$ & 1\\[.4cm]
      $L_R = \bpm n_R\\e_R\epm$  &
        $\big({\bf 1}, {\bf 1}, {\bf 2},  -\frac12\big)$ & $\frac32$\\[.4cm]
      $n_L$   & $\big({\bf 1}, {\bf 1}, {\bf 1}, 0\big)$ & 2\\[.2cm]
      $\nu_R$ & $\big({\bf 1}, {\bf 1}, {\bf 1}, 0\big)$ & 1
    \end{tabular}\hspace*{.5cm}
    \begin{tabular}{ccc}
      Fields & Repr.&$U(1)_S$\\\hline\hline\\[-.4cm]
      $\phi = \bpm \phi_1^0&\phi_2^+\\ \phi_1^-&\phi^0_2 \epm$ &
         $\big({\bf 1}, {\bf 2}, {\bf 2}^*, 0\big)$ & $-\frac12$\\[.4cm]
      $\chi_L = \bpm\chi_L^+ \\\chi_L^0\epm$ &
         $\big({\bf 1}, {\bf 2}, {\bf 1}, \frac12\big)$ & 0 \\[.4cm]
      $\chi_R = \bpm\chi_R^+ \\\chi_R^0\epm$ &
         $\big({\bf 1}, {\bf 1}, {\bf 2}, \frac12\big)$ &$\frac12$\\[.4cm]\hline
      $G_\mu$     & $\big({\bf 8}, {\bf 1}, {\bf 1}, 0\big)$ &0\\[.2cm]
      $W_{L \mu}$ & $\big({\bf 1}, {\bf 3}, {\bf 1}, 0\big)$ &0\\[.2cm]
      $W_{R \mu}$ & $\big({\bf 1}, {\bf 1}, {\bf 3}, 0\big)$ &0\\[.2cm]
      $B_\mu$     & $\big({\bf 1}, {\bf 1}, {\bf 1}, 0\big)$ &0\\[.2cm]
    \end{tabular}
    \caption{ALRSM particle content, given together with the representation of
    each field under $SU(3)_c\times SU(2)_L\times SU(2)_{R'} \times U(1)_{B-L}$
    (second column) and the $U(1)_S$ quantum numbers (third column). We consider
    the matter sector (left panel), the gauge sector (lower right panel) and the
    Higgs sector (upper right panel) separately.}
    \label{tab:content}
  \end{center}
\end{table}

Pairing the fields presented in eq.~\eqref{eq:331_fields} into $SU(3)_c \times
SU(2)_L\times SU(2)_{R'}\times U(1)_{B-L}$ multiplets yields phenomenological
issues for the neutrino sector, as the lightest neutrinos get
masses of the order of the up quark mass~\cite{Frank:2004vg}. This can be cured
by adding an $E_6$ singlet scotino
$n_L$ to the field content, together with a pair of (heavy) ${\bf 27} + {\bf
\overline{27}}$ Higgs fields. As a consequence, the exotic $E$ and $N$ fermions
become much heavier and can be phenomenologically ignored. The resulting
fermionic content of the model is presented in the left panel of
table~\ref{tab:content}, together with the representations under the model gauge
group and the associated $U(1)_S$ quantum numbers. The electric charge of the
different fields can be obtained through a generalised Gell-Mann-Nishijima
relation $Q = T_{3R} + T_{3L} + Y_{B-L}$, which subsequently explains the
unconventional $B-L$ charges.

In order to recover the electroweak symmetry group, the gauge and global
symmetry $SU(2)_{R'} \times U(1)_{B-L} \times U(1)_S$ is first broken down to
the hypercharge $U(1)_Y$ while preserving the generalised lepton number $L$.
This is achieved through an $SU(2)_{R'}$ doublet of scalar fields $\chi_R$
charged under $U(1)_S$. While we introduce an $SU(2)_L$ counterpart $\chi_L$ to
maintain the left-right symmetry, the latter is in contrast blind to the global
$U(1)_S$ symmetry. The electroweak symmetry is then broken down to
electromagnetism by means of a bidoublet of Higgs fields charged under both
$SU(2)_L$ and $SU(2)_{R'}$, but with no $B-L$ quantum numbers. We refer to the
right panel of table~\ref{tab:content} for details on the gauge and Higgs
sector of the ALRSM.

The model Lagrangian includes, on top of standard gauge-invariant kinetic terms
for all fields, a Yukawa interaction Lagrangian $\lag_{\rm Y}$ and a scalar
potential $V_{\rm H}$. The most general Yukawa Lagrangian allowed by the gauge
and the global $U(1)_S$ symmetries is given by
\be\label{eq:yuk}
  \lag_{\rm Y} = \bar Q_L {\bf \hat Y}^u \hat\phi^\dag Q_R
    - \bar Q_L {\bf \hat Y}^d \chi_L d_R
    - \bar Q_R {\bf \hat Y}^{d'}\chi_R d'_L
    - \bar L_L {\bf \hat Y}^e \phi L_R
    + \bar L_L {\bf \hat Y}^\nu \hat\chi_L^\dag \nu_R
    + \bar L_R {\bf \hat Y}^n \hat\chi_R^\dag n_L + {\rm h.c.} \ ,
\ee
where all flavour indices have been omitted for clarity so that the Yukawa
couplings ${\bf \hat Y}$ are $3 \times 3$ matrices in the flavour space,
 and where the hatted quantities refer to the duals of the scalar fields
$\hat\phi=\sigma_2\phi\sigma_2$ and $\hat\chi_{L,R}= i \sigma_2 \chi_{L,R}$ (with
$\sigma_2$ being the second Pauli matrix).
The most general Higgs potential $V_{\rm H}$ preserving the left-right symmetry
is given, following standard conventions~\cite{Borah:2010zq}, by
\be\bsp
  V_{\rm H} = &
   -\mu_1^2 {\rm Tr} \big[\phi^\dag \phi\big]
   -\mu_2^2 \big[\chi_L^\dag \chi_L + \chi_R^\dag \chi_R\big]
   + \lambda_1 \big({\rm Tr}\big[\phi^\dag \phi\big]\big)^2
   + \lambda_2\ (\phi\!\cdot\!\hat\phi)\ (\hat\phi^\dag\!\cdot\!\phi^\dag)
\\&
   + \lambda_3 \Big[\big(\chi_L^\dag \chi_L\big)^2 +
         \big(\chi_R^\dag\chi_R\big)^2\Big]
   + 2 \lambda_4\ \big(\chi_L^\dag \chi_L\big)\ \big(\chi_R^\dag\chi_R\big)
   + 2 \alpha_1 {\rm Tr} \big[\phi^\dag \phi\big]
          \big[\chi_L^\dag \chi_L + \chi_R^\dag \chi_R\big]
\\&
   + 2 \alpha_2 \big[ \big(\chi_L^\dag \phi\big) \big(\chi_L\phi^\dag\big) +
          \big(\phi^\dag \chi_R^\dag\big)\ \big(\phi\chi_R\big)\big]
   + 2 \alpha_3 \big[ \big(\chi_L^\dag \hat\phi^\dag\big)\
          \big(\chi_L\hat\phi\big) + \big(\hat\phi\chi_R^\dag\big)\
          \big(\hat\phi^\dag\chi_R\big)\big]
\\&
   + \kappa \big[\chi_L^\dag \phi \chi_R + \chi_R^\dag\phi^\dag\chi_L\big] \ ,
\esp\label{eq:Hpot}\ee
and contains bilinear ($\mu$), trilinear ($\kappa$) and quartic 
($\lambda$, $\alpha$) contributions. In the above expression, the dot to the
$SU(2)$-invariant product.

After the breaking of the left-right symmetry down to electromagnetism, the
neutral components of the scalar fields acquire non-vanishing vacuum expectation
values (vevs),
\be
  \langle \phi  \rangle = \frac{1}{\sqrt{2}}\bpm 0&0\\0 & k \epm\ , \qquad
  \langle \chi_L\rangle = \frac{1}{\sqrt{2}}\bpm 0\\ \vL \epm\ , \qquad
  \langle \chi_R\rangle = \frac{1}{\sqrt{2}}\bpm 0\\ \vR \epm\ ,
\label{eq:symbreak}\ee
with the exception of $\phi^0_1$, which is protected by the conservation of the
generalised lepton number that also forbids mixing between the SM $d$ and
exotic $d^\prime$ quarks.
Moreover, all scalar fields with the same electric
charge mix. Expressing the complex neutral scalar fields in terms of their real
degrees of freedom,
\be\bsp
 \renewcommand{\arraystretch}{1.3}
  \phi_1^0 =&\ \frac{1}{\sqrt{2}}
     \Big[ \Re\{\phi^0_1\} + i\ \Im\{\phi^0_1\}\Big]\ ,
\\
  \phi_2^0 =&\ \frac{1}{\sqrt{2}}
     \Big[ k + \Re\{\phi^0_2\} + i\ \Im\{\phi^0_2\}\Big] \ ,
\\
  \chi_{L,R}^0 =&\ \frac{1}{\sqrt{2}}
     \Big[ v_{\scriptscriptstyle L,R} + \Re\{\chi^0_{L,R}\} +
     i\ \Im\{\chi^0_{L,R}\}\Big]\ ,
\esp\ee
we can write the mixing relations involving the massive $CP$-even Higgs bosons
$H_i^0$ (with $i=0,1,2,3$), the massive $CP$-odd Higgs bosons $A_i^0$ (with
$i=1,2$) and the two massless Goldstone bosons $G_1^0$ and $G_2^0$ that give
rise to the longitudinal degrees of freedom of the $Z$ and $Z'$ bosons, as
\be
 \bpm \Im\{\phi_1^0\}\\\Im\{\phi_2^0\}\\\Im\{\chi_L^0\}\\\Im\{\chi_R^0\}\epm =
 \bpm 1 & 0 & 0 & 0\\ 0&&& \\ 0&& U_{3\times 3}^{\rm A} & \\0&&& \\ \epm
 \bpm A_1^0\\ G_1^0 \\ G_2^0\\A_2^0\epm \qquad\text{and} \qquad
 \bpm \Re\{\phi_1^0\}\\\Re\{\phi_2^0\}\\\Re\{\chi_L^0\}\\\Re\{\chi_R^0\}\epm =
 \bpm 1 & 0 & 0 & 0\\ 0&&& \\ 0&& U_{3\times 3}^{\rm H} & \\0&&& \\ \epm
  \bpm H_1^0\\ H_0^0 \\ H_2^0\\H_3^0\epm   \ .
\label{eq:nh_mix}\ee
The $\phi_1^0$ field has been prevented from any mixing by virtue of the
conservation of the generalised lepton number, and we refer to
appendix~\ref{app:higgs} for the expressions of the $3\times 3$ Higgs mixing
matrices $U_{3\times 3}^{\rm A}$ and $U_{3\times 3}^{\rm H}$, as
well as for those of the six Higgs-boson masses.
In the charged sector, the $\phi_1^\pm$, $\phi_2^\pm$, $\chi_L^\pm$ and
$\chi_R^\pm$ fields mix into two physical massive charged Higgs bosons $H_1^\pm$
and $H_2^\pm$, as well as  two massless Goldstone bosons $G_1^\pm$ and $G_2^\pm$
that are absorbed by the $W$ and $W'$ gauge bosons,
\be\bsp
 \bpm \phi_2^\pm\\\chi_L^\pm\epm =
 \bpm \cos\beta & \sin\beta\\ -\sin\beta & \cos\beta\epm
 \bpm H_1^\pm\\G_1^\pm\epm \ , \ \
 \bpm \phi_1^\pm\\\chi_R^\pm\epm =
 \bpm \cos\zeta & \sin\zeta\\ -\sin\zeta & \cos\zeta\epm
 \bpm H_2^\pm\\G_2^\pm\epm \ ,
\esp\label{eq:ch_mix}\ee
with
\be
  \tan\beta = \frac{k}{\vL} \qquad\text{and}\qquad \tan\zeta = \frac{k}{\vR} \ .
\ee
We refer again to appendix~\ref{app:higgs} for the explicit expressions of the masses of
the physical states in terms of other model parameters.

By definition, the breaking of the left-right symmetry generates masses for the
model gauge bosons and induces their mixing (from the Higgs-boson kinetic
terms). The charged $W=W_L$ and $W'=W_R$ bosons do not mix as $\langle\phi_1^0
\rangle = 0$, and their masses are given by
\be
  M_W    = \frac12 \gL \sqrt{k^2+\vL^2} \equiv \frac12 \gL v
 \qquad\text{and}\qquad
  M_{W'} = \frac12 \gR \sqrt{k^2+\vR^2} \equiv \frac12 \gR v' \ .
\label{eq:mw_mwp}\ee
In the neutral sector, the gauge boson squared mass matrix is written, in the
$(B_\mu, W_{L\mu}^3, W_{R\mu}^3)$ basis, as
\be
  ({\cal M}^0_V)^2 = \frac14 \bpm
    \gBL^2\ (\vL^2+\vR^2)  & -\gBL\ \gL\ \vL^2    & -\gBL\ \gR\ \vR^2\\
   -\gBL\ \gL\ \vL^2       &  \gL^2\ v^2          & -\gL\ \gR\ k^2\\
   -\gBL\ \gR\ \vR^2       & -\gL\ \gR\ k^2       &  \gR^2\ v^{\prime 2}
  \epm \ .
\ee
It can be diagonalised through three rotations that mix the $B$, $W_L^3$ and
$W_R^3$ bosons into the massless photon $A$ and massive $Z$ and $Z'$ states,
\renewcommand{\arraystretch}{1.}
\be
  \bpm B_\mu\\ W_{L\mu}^3\\ W_{R\mu}^3\epm = 
  \bpm c_{\phw} & 0 & -s_{\phw}\\ 0 & 1 & 0\\ s_{\phw} & 0 & c_{\phw} \epm
  \bpm c_{\tw} & -s_{\tw} & 0\\ s_{\tw} & c_{\tw} & 0\\ 0 & 0 & 1 \epm
  \bpm 1 & 0 & 0\\ 0 & c_{\zw} & -s_{\zw}\\ 0 & s_{\zw} & c_{\zw} \epm
  \bpm A_\mu\\ Z_\mu\\ Z^\prime_\mu\epm  \ ,
\ee
where $s_i$ and $c_i$ respectively denote the sine and cosine of the angle $i$.
The $\phw$-rotation mixes the $B$ and $W_R^3$ bosons into the hypercharge boson
$B'$ as generated by the breaking of $SU(2)_{R'}\times U_{B-L}$ into to the
hypercharge group $U(1)_Y$. The $\tw$-rotation denotes the usual electroweak
mixing, and the $\zw$-rotation is related to the strongly constrained $Z$/$Z'$
mixing. The various mixing angles are defined by
\be\bsp
 & s_{\phw} = \frac{\gBL}{\sqrt{\gBL^2+\gR^2}} = \frac{\gY}{\gR}
   \qquad\text{and}\qquad
   s_{\tw}  = \frac{\gY}{\sqrt{\gL^2+\gY^2}} = \frac{e}{\gL} \ ,
 \\&
  \tan(2\zw)=\frac{2 c_{\phw} c_{\tw} \gL \gR (c_{\phw}^2 k^2-s_{\phw}^2\vL^2)}
     {-(\gL^2 - c_{\phw}^2 c_{\tw}^2 \gR^2) c_{\phw}^2 k^2 -
         (\gL^2 - c_{\tw}^2 \gBL^2 s_{\phw}^2) c_{\phw}^2 \vL^2 +
          c_{\tw}^2 \gR^2 \vR^2} \ ,
\esp\label{eq:ewmix}\ee
where $\gY$ and $e$ denote the hypercharge and electromagnetic coupling
constant respectively. Neglecting the $Z$/$Z'$ mixing, the $Z$ and $Z'$ boson
masses are given by
\be
  M_{Z} =  \frac{\gL}{2 c_{\tw}} \ v
  \qquad\text{and}\qquad
  M_{Z'} = \frac12 \sqrt{\gBL^2 s_{\phw}^2 \vL^2 +
     \frac{\gR^2 (c_{\phw}^4 k^2 + \vR^2)}{c_{\phw}^2}} \ .
\label{eq:mz_mzp}\ee

The breaking of the gauge symmetry furthermore generates masses and mixings in
the fermion sector. The masses of the up-type quark and charged leptons are
controlled by the vev $k$ of the Higgs bidoublet, whereas the masses of the
neutrinos and the down-type quarks arise from the vev $\vL$ of the $\chi_L$
Higgs triplet. The scale of the exotic fermion masses is in contrast solely
induced by the vev $\vR$ of the $\chi_R$ triplet. Similarly
to what is achieved in the LRSM, all fermion mixing are conveniently absorbed into
two CKM ($V_{\rm CKM}$ and $V_{\rm CKM'}$) and two PMNS ($V_{\rm PMNS}$ and
$V_{\rm PMNS'}$) rotations,
\be
  d_L  \to V_{\rm CKM} d_L \ , \quad
  \nu_L\to V_{\rm PMNS} d_L \ , \quad
  d'_R \to V_{\rm CKM'} d'_R\ , \quad
  n_R  \to V_{\rm PMNS'} n_R \ .
\ee
We refer to appendix~\ref{app:ferms} for additional details on the generation of
the fermion masses, and their explicit expression in terms of the other model
free parameters.

Finally, we supplement the model Lagrangian by the effective couplings
$a_{\rm H}^g$ and $a_{\rm H}^a$ of the SM Higgs boson to gluons and photons,
\be
 \lag_{\rm eff} = -\frac14 a_{\rm H}^g H_0^0 G_{\mu\nu}^a G^{\mu\nu}_a
   - \frac14 a_{\rm H}^a H_0^0 F_{\mu\nu} F^{\mu\nu} \ ,
\ee
where $G_{\mu\nu}^a$ and $F_{\mu\nu}$ respectively denote the gluon and photon
field strength tensors.

\section{Computational setup}
\label{sec:scan}

To perform our analysis of the cosmology and collider phenomenology of the
ALRSM, we have implemented the model presented in section~\ref{sec:ALRSMmodel}
into \fr\ (version 2.3.35)~\cite{Alloul:2013bka}. Whereas an implementation was
already publicly available for many years~\cite{Ashry:2013loa, AshryThesis}, we found several
issues with the latter that justified the development of a new implementation
from scratch.
First, the Goldstone sector is incorrectly implemented in the existing
implementation, which could yield wrong predictions when jointly used with a
tool handling computations in Feynman gauge by default (like \mo~\cite{
Belanger:2018mqt}). Secondly, all scalar fields are doubly-declared (\ie\ both
under their standard and dual form), the implementation is only partly relying
on \fr\ built-in functions to treat index contractions and covariant
derivatives, and the declaration of the model parameters relies particularly heavily
on the existence of an unnecessary large amount of temporary
intermediate abbreviations. This consequently renders the implementation hard
to verify and understand. Moreover, the electroweak sector is defined by five
independent parameters instead of three. Thirdly, the existing implementation
enforces the unnecessary equality $\gL=\gR$, that is justified neither theoretically nor
phenomenologically. Relaxing this constraint would have required to modify all
relations relevant for the gauge and Higgs boson masses and mixings (see
section~\ref{sec:ALRSMmodel} and appendix~\ref{app:higgs}), which would have
been quite a complex task given the heavy handling of the model
parameters. Finally, the original implementation has also the $V_{\rm CKM} =
V_{\rm CKM'}$ and $V_{\rm PMNS}=V_{\rm PMNS'}$ equalities built in, which is
again not justified (see appendix~\ref{app:ferms}). For all
those reasons, we decided on designing a fresh, more general, implementation,
that is also publicly released on the \fr\ model database\footnote{See
\href{http://feynrules.irmp.ucl.ac.be/wiki/ALRM_general}
{http://feynrules.irmp.ucl.ac.be/wiki/ALRM\_general}.}.
In order to facilitate the usage of our \fr\ implementation, we document it
further in appendix~\ref{app:fr}, where we provide information on the new
physics mass-eigenstates supplementing the SM field content, the free  model  
parameters and their relation to all the other (internal) parameters.

We have then made use of \fr\ to generate \ch~\cite{Belyaev:2012qa} model files
and a UFO~\cite{Degrande:2011ua} version of the model~\cite{Christensen:2009jx},
so that we could employ \mo\ (version 5.0.8)~\cite{Belanger:2018mqt} for the
computation of the predictions relevant for our dark matter study, and \mg\
(version 2.6.4)~\cite{Alwall:2014hca} for generating the hard-scattering event
samples necessary for our collider study. These events, obtained by convoluting
the hard-scattering matrix elements with the leading-order set of NNPDF~2.3
parton densities~\cite{Ball:2012cx}, are subsequently matched with the \py\
(version 8.243)~\cite{Sjostrand:2014zea} parton showering and hadronisation
algorithms, and we simulate the typical response of an LHC detector by means of
the \del~\cite{deFavereau:2013fsa} programme (version 3.4.2) that internally
relies on the anti-$k_T$ algorithm~\cite{Cacciari:2008gp} as implemented into
\fj~\cite{Cacciari:2011ma} (version 3.3.2) for event reconstruction. We have
employed \ma~\cite{Conte:2012fm} (version 1.8.23) for the collider
analysis of section~\ref{sec:collider}. Moreover, we have additionally used the
generated UFO model with \mdm~\cite{Ambrogi:2018jqj} to independently verify the
results obtained with \mo\, , in particular for what concerns gauge invariance.

\begin{table}
  \centering
  \setlength\tabcolsep{8pt}
  \renewcommand{\arraystretch}{1.3}
  \begin{tabular}{cc|cc}
    Parameter  & Scanned range & Parameter      & Scanned range \\
    \hline
    $\tan\beta$ & $[0.7, 50]$      & $m_{n_1}$    & $[10, 2000]$ GeV\\
    $\gR$       & $[0.37, 0.768]$  & $m_{n_2}$  & $[10, 2000]$ GeV\\
    $v'$        & $[6.5, 13]$~ TeV & $m_{n_3}$ & $[10, 2000]$ GeV\\
    \hline
    $\lambda_2$ & 0.               & $m_{d'}$ & $[500, 2000]$ GeV \\
    $\lambda_3$ & $[0.01, 0.09]$   & $m_{s'}$ & $[m_{d'}, 2500]$ GeV \\
    $\kappa$    & $[-50, -1]$~GeV  & $m_{b'}$ & $[m_{s'}, 3000]$ GeV \\
    $\alpha_1=\alpha_2=\alpha_3$ & $[ 0.01, 0.5]$  \\
  \end{tabular}
  \caption{Ranges where the new parameters defining the new physics sector of
    the model are allowed to vary.}
  \label{tab:scan_lim}
\end{table}

In addition, we have relied on \hb\ (version 4.3.1)~\cite{Bechtle:2008jh} and
\hs\ (version 1.4.0)~\cite{Bechtle:2013xfa} to verify the compatibility of the
ALRSM Higgs sector with data, with the $H_0^0$ field being associated with the SM
Higgs boson. We have used the \pysl\ package~\cite{Buckley:2013jua} to
read the input values for the model parameters that we encode under the SLHA
format~\cite{Skands:2003cj}, and to integrate the various employed programmes
into a single framework. Using our interfacing, we performed a random scan
of the model parameter space following the Metropolis-Hastings technique. We
have fixed the SM parameters to their Particle Data Group (PDG) values~\cite{
Tanabashi:2018oca}, chosen the $V_{\rm CKM'}$ and $V_{\rm PMNS'}$ matrices to be equal to
their SM counterparts, and varied the remaining 15 parameters as described in
table~\ref{tab:scan_lim}.

The $SU(2)_{R'}$ coupling $\gR$ is allowed to vary within the $[0.37, 0.768]$
window. The lower bound originates from the $\gR/\gL$ ratio that is
theoretically constrained to be larger than $\tan\tw$~\cite{Dev:2016dja},
whereas the upper bound is phenomenological. In practice, $\gR$ can indeed vary
all up to the perturbative limit of $\gR = \sqrt{4\pi}$. However, imposing an
upper bound on $\gR$ that is 4--5 times smaller guarantees scenarios that are
viable with respect to LHC limits~\cite{Sirunyan:2018rlj,Sirunyan:2018exx,
Aaboud:2017buh,Aaboud:2017yvp} and that feature at
least one light extra gauge boson (see section~\ref{sec:gaugebosons}). The same
light-spectrum considerations has lead to our choices for the values of the
$\tan\beta$ and $v'$ parameters, with the additional constrains stemming from
the expectation that the $SU(2)_{R'}$ symmetry has to be broken in the multi-TeV regime
and that the $Z/Z'$ mixing must be negligibly small.

The ranges and configuration adopted for the parameters of the Higgs sector are
driven by the Higgs potential minimisation conditions of eqs.~\eqref{eq:cstr2}
and \eqref{eq:minpot}, as well as by the above-mentioned LHC constraints on the
$Z'$-boson mass, and by the requirement that the lightest charged Higgs boson is not
tachyonic. It turns out that all phenomenologically acceptable scenarios feature
$\alpha_1\sim\alpha_2 = \alpha_3$ and $\lambda_2=0$, so that we set for
simplicity
\be
  \lambda_2 = 0\quad\text{and}\quad \alpha_1 = \alpha_2 =\alpha_3 \ .
\ee
Moreover, $\lambda_3$ has to be small and we recall that $\kappa$ has to be
negative (see appendix~\ref{app:higgs}). Finally, the exotic quarks and scotino
masses are not restricted and we allow them to vary mostly freely, with a
phenomenological upper bound allowing  them to be not too heavy.

\section{Gauge boson mass constraints}
\label{sec:gaugebosons}

\begin{figure}
  \centering
  \includegraphics[width=0.49\columnwidth]{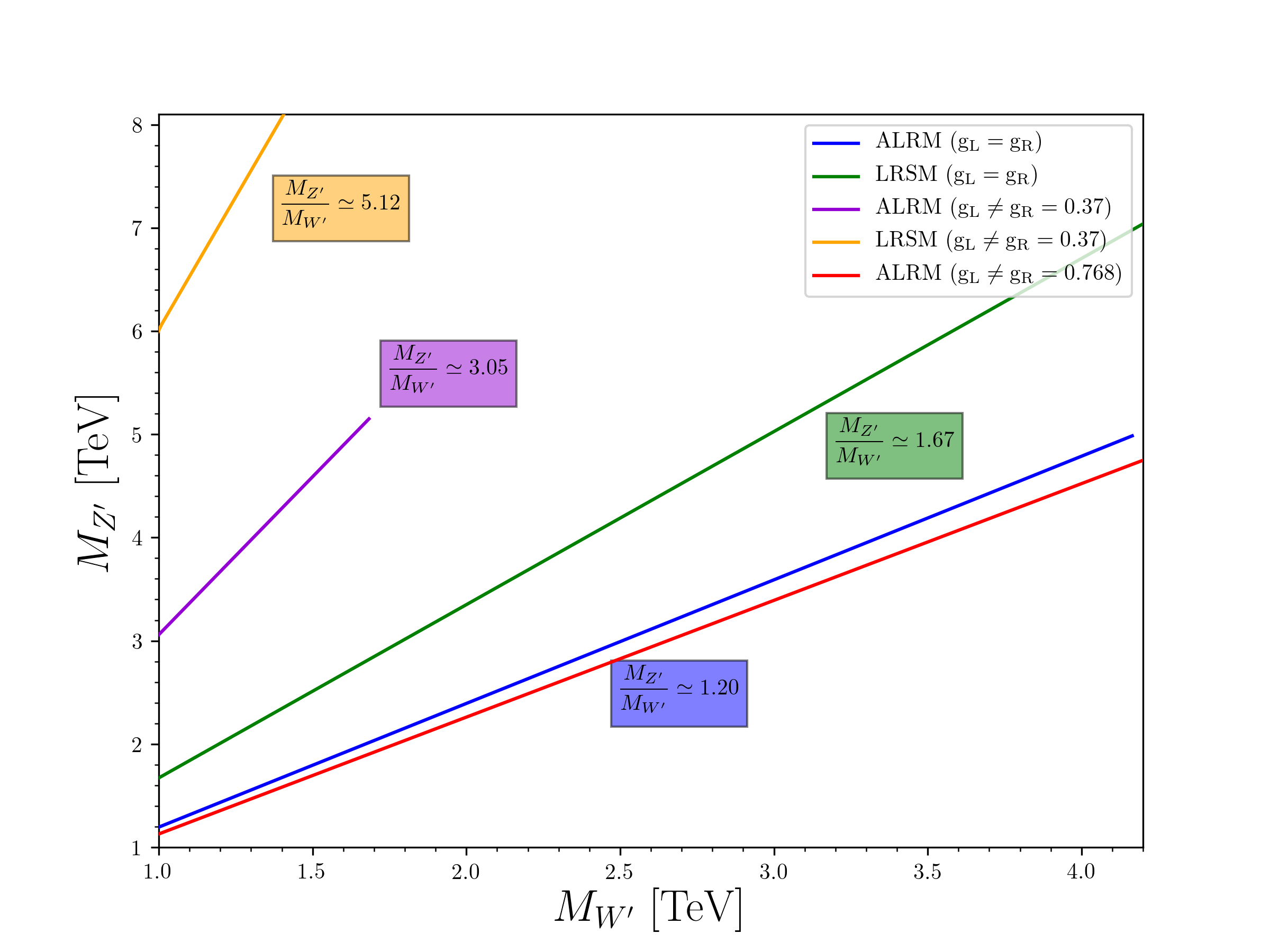}
  \includegraphics[width=0.49\columnwidth]{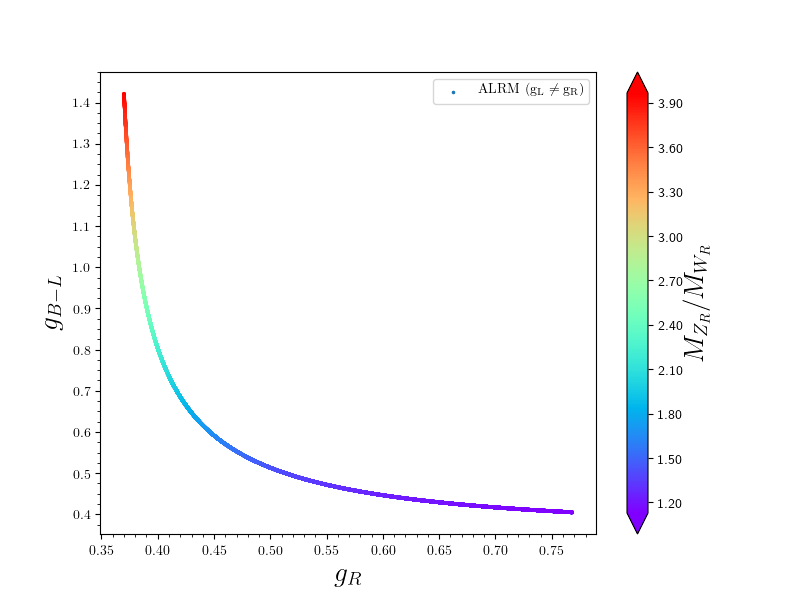}
  \includegraphics[width=0.49\columnwidth]{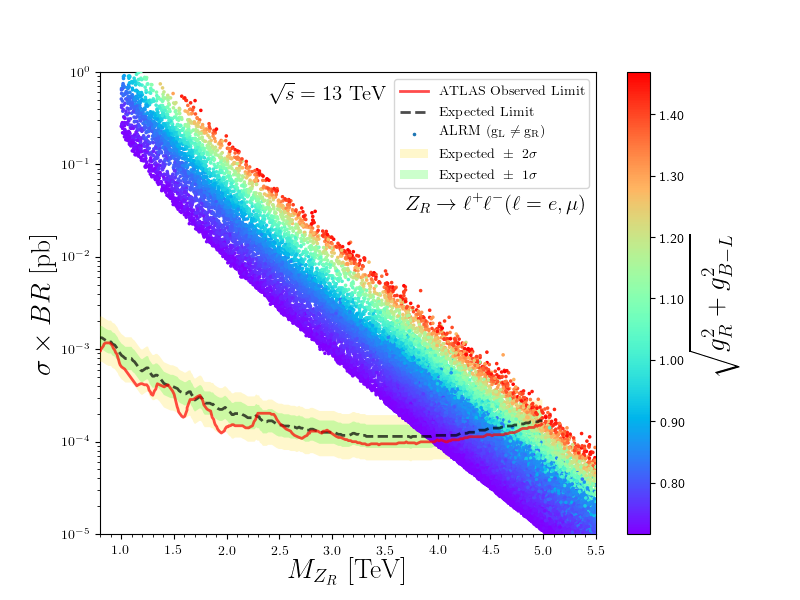}
  \includegraphics[width=0.49\columnwidth]{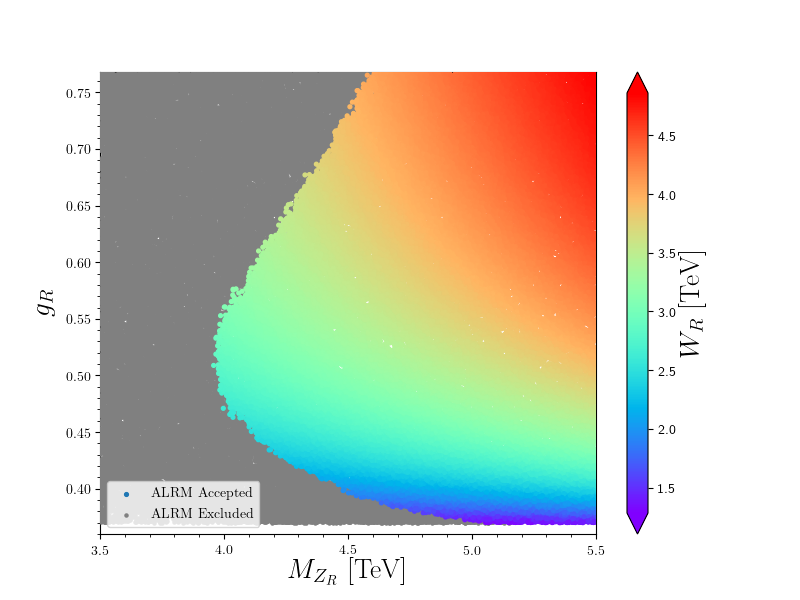}
  \caption{Properties of the gauge sector for the ALRSM scenarios featuring a
    Higgs sector compatible with data. We emphasise the relations between the
    $W'$ and $Z'$ boson masses with the gauge couplings and also investigate the
    LHC constraints on the mass of the $Z'$ boson.}
  \label{fig:ZpLimits}
\end{figure}

 Following the methodolgy described in the previous
section, we scan the parameter space imposing constraints on the properties of the Higgs sector so that the
$H_0^0$ scalar boson is SM-like and has features agreeing with experimental data. In this
section, we analyse the properties of the gauge sector for all scenarios
accepted in our scanning procedure.

In the upper left and right panels of figure~\ref{fig:ZpLimits}, we depict the
relations between the masses of the extra gauge bosons $M_{Z'}$ and $M_{W'}$ and
the ALRSM coupling constants $\gL$, $\gBL$ and $\gR$. We observe, in the upper
left panel of the figure, that in the ALRSM the ratio of the neutral to the
charged extra boson masses ranges from about 1.20 for a maximal $\gR$ value of
0.768 (light green line) to about 3.05 for a minimal setup defined by $\gR=0.37$
(purple line). The left-right symmetric case $\gL=\gR\approx 0.64$ is also
indicated (dark blue line). This shows that a large variety of splittings can be
realised for gauge boson masses lying in the 1--5 TeV range. Equivalently, both
compressed spectra in which the $Z'$-boson is only 20\% heavier than the $W'$-boson and more
split spectra in which the $Z'$-boson is more than about 3 times heavier than the $W'$-boson
are allowed by Higgs data, and this for a large set of $W'$-boson masses lying
in the 1--4~TeV range. We compare those findings with predictions relevant for
the usual LRSM for similar $\gR$ values (dark green and orange lines for $\gL =
\gR$ and $\gR=0.37$ respectively). It turns out that the $M_{Z'}/M_{W'}$ ratio
is lower in the ALRSM than in the LRSM for a given $\gR$ value, \ie\ the ALRSM
gauge boson spectrum is more compressed than in the standard LRSM for a given
$SU(2)_R$ coupling constant value. In the upper right panel of
figure~\ref{fig:ZpLimits}, we study the dependence of this mass ratio on the
$\gBL$ an $\gR$ coupling constants. The latter two couplings are related to the
hypercharge coupling,
\be
\frac{1}{\gY^2} = \frac{1}{\gR^2} + \frac{1}{\gBL^2} \ ,
\ee
so that large $\gR$ values are always associated with low $\gBL$ values and {\it
vice versa}. In typical scenarios, the hierarchy $\vL \ll k \ll \vR$ is
fulfilled as $\vL$ is small (which is also favoured by constraints originating
from the $\rho$ parameter~\cite{Arhrib:2011uy}), $k$ drives the electroweak
vacuum and is of ${\cal O}(100)$~GeV, and $\vR$ is related to the breaking of
the $SU(2)_{R'}$ symmetry and is thus larger. Therefore, eqs.~\eqref{eq:mw_mwp},
\eqref{eq:ewmix} and \eqref{eq:mz_mzp} yield
\be
  \frac{M_{Z'}}{M_{W'}} \approx \frac{1}{c_{\phw}} = \frac{\gBL}{\gY}\ .
\ee
When $\gR$ is larger, $\gBL$ is smaller and $c_{\phw}$ is consequently larger.
Smaller $M_{Z'}/M_{W'}$ ratios are thus expected. Conversely, with increasing
values of $\gBL$, $c_{\phw}$ and $\gR$ become smaller so that the $M_{Z'}/M_{W'}$
ratio increases. In those case, the $W'$ boson can become up to about three times
lighter than the $Z'$-boson (see the upper left panel of the figure). This
feature has profound consequences on the possible existence of light ALRSM $W'$
bosons allowed by data.

The $W'$-boson does not indeed couple to pairs of ordinary SM fermions, but
instead couples to a SM up-type quark and an exotic down-type quark $d'$, or an
electron and a scotino. It can consequently not be directly produced at
colliders and all LHC bounds on an additional $W'$ boson originating from dijet
and dileptonic resonance searches are automatically evaded~\cite{%
Sirunyan:2018rlj,Sirunyan:2018exx,Aaboud:2017buh,Aaboud:2017yvp}. Only the
neutral ALRSM $Z'$-boson can potentially be searched for through standard extra
gauge boson LHC analyses, as it is allowed to couple to pairs of SM fermions. We
evaluate the resulting bounds in the lower left panel of
figure~\ref{fig:ZpLimits} in which
we consider the most constraining limits originating from the cleaner
searches in the dilepton mode. For each benchmark scenario selected by our
scanning procedure, we evaluate the $Z'$-boson production cross section,
including the branching ratio associated with a $Z'\to e^+ e^-$ or $\mu^+\mu^-$
decay, and compare our predictions to the bounds arising from the ATLAS search
of ref.~\cite{Aaboud:2017buh}. The spread in cross section obtained for a given
$Z'$ mass stems from the different values of the strength of the
$Z'$-boson fermionic couplings, which we estimate by $\sqrt{\gR^2 + \gBL^2}$ and
which is represented through the colour map in the figure. For the smallest coupling
values, $Z'$ bosons as light as
4~TeV are allowed by data, whilst when the coupling strength gets larger, the
limits can be pushed up to 5~TeV\footnote{Whilst in the large coupling case,
the $Z'$ width over mass ratio can reach 10\%, we have verified that our
approximation in which we neglect the interferences of the signal with the SM
dilepton continuum was reasonably satisfactory.}.

As previously mentioned and visible from the upper left panel of
figure~\ref{fig:ZpLimits}, the $W'$- and $Z'$-bosons can feature a very split
spectrum so that a 4-5~TeV $Z'$ boson can coexist with a 1--2 TeV $W'$-boson.
This feature is illustrated in the lower right panel of the figure in which we
present, for each scenario satisfying the LHC $Z'$ bounds (the excluded
benchmarks being shown in grey), the corresponding value of the $\gR$ coupling.
The latter dictates the $W'$-boson mass value, as given by eq.~\eqref{eq:mw_mwp}
which we also represent through the colour map. For the lowest $\gR$ values
allowed in the scan, the additional gauge boson splitting is expected to be the
largest (see the upper left panel of figure~\ref{fig:ZpLimits}), so that viable
scenarios featuring a $W'$ boson as light as 1--2~TeV and a $Z'$-boson not excluded by present searches are found. The considered $Z'$ bounds are expected to slightly improve by
about 20\% during the high-luminosity operation phase of the
LHC~\cite{ATL-PHYS-PUB-2018-044}, which does not challenge the existence of
light $W'$ bosons (see the lower right panel of figure~\ref{fig:ZpLimits}).
The lightest options for the $W'$ boson correspond to scenarios
featuring the smallest $\gR$ value theoretically allowed ($\gR\sim 0.37$), the
$Z'$-boson being in this case constrained to lie above roughly 5~TeV. Viable
scenarios in which the $Z'$-boson is lighter, with $M_{Z'}\approx 4$~TeV, are
also allowed by data. In that configuration, the $U(1)_{B-L}$ and $SU(2)_{R'}$
coupling constant are of a similar magnitude, $\gR\approx\gBL\sim 0.5$ (see the
upper right panel of figure~\ref{fig:ZpLimits}), and the $W'$/$Z'$ boson
splitting is smaller ($M_{W'}\approx 3$~TeV). Our results also show that the
largest $\gR$ values correspond to the heaviest scenarios, being thus
disfavoured to be observed at current colliders. This motivates the upper
bound set on $\gR$ in our scan (see section~\ref{sec:scan}).

\section{Dark matter}
\label{sec:dm}

\begin{figure}
  \includegraphics[width=.49\columnwidth]{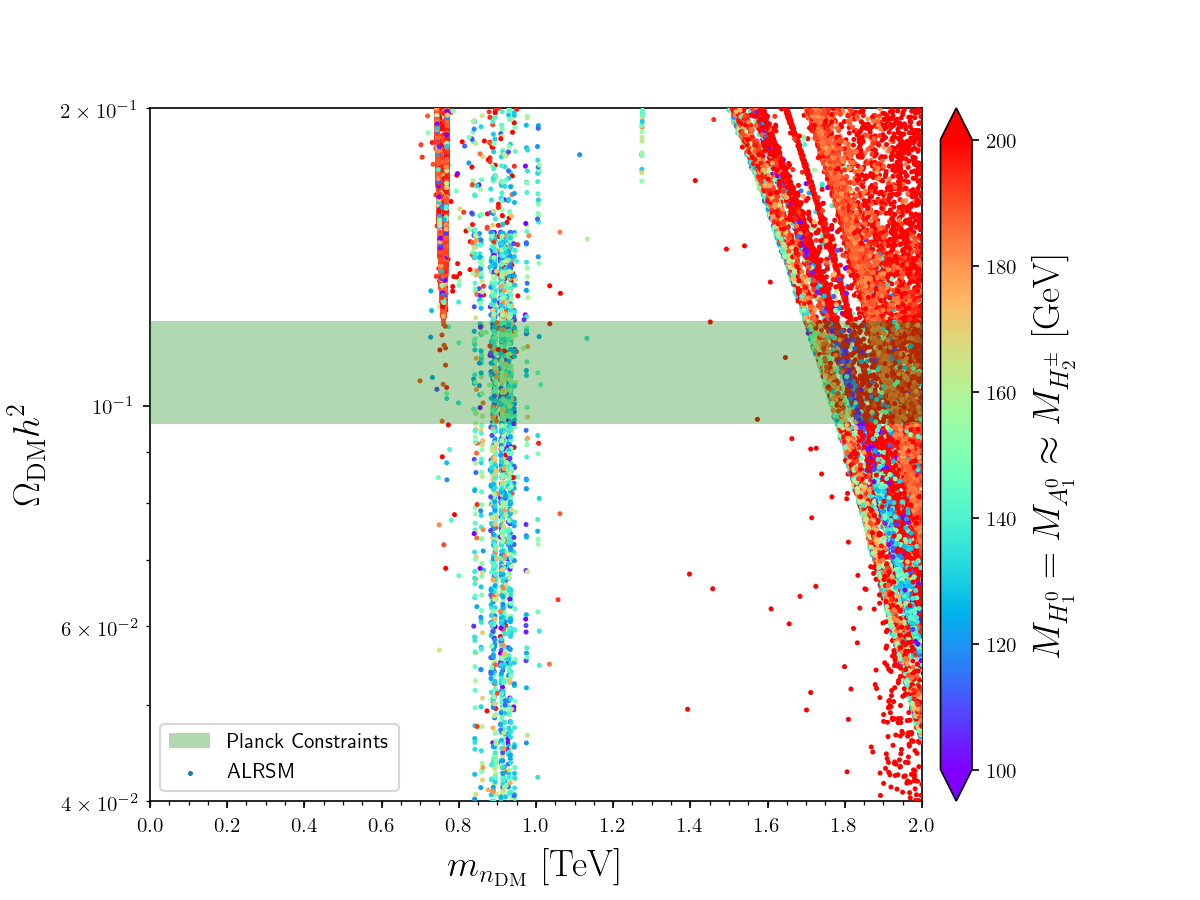}
  \includegraphics[width=.49\columnwidth]{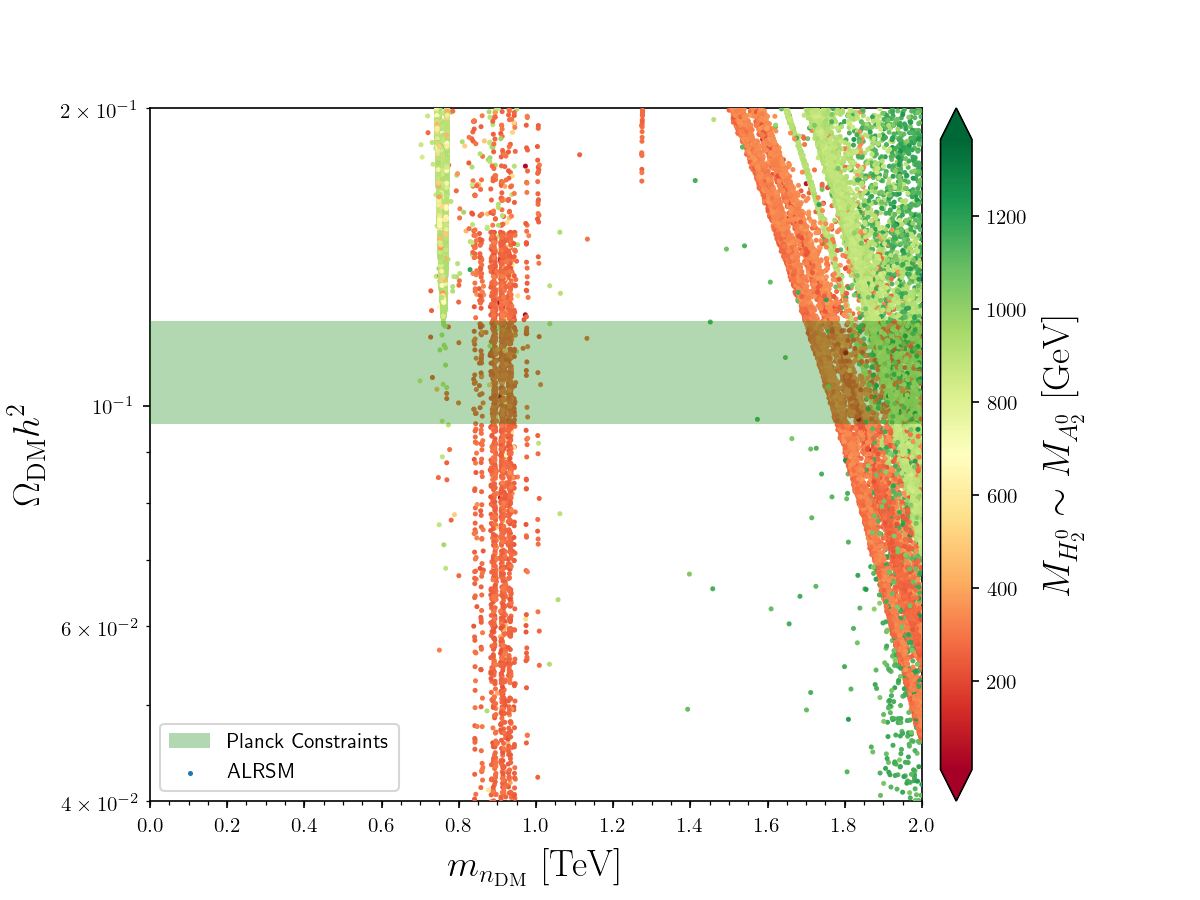}
  \includegraphics[width=.49\columnwidth]{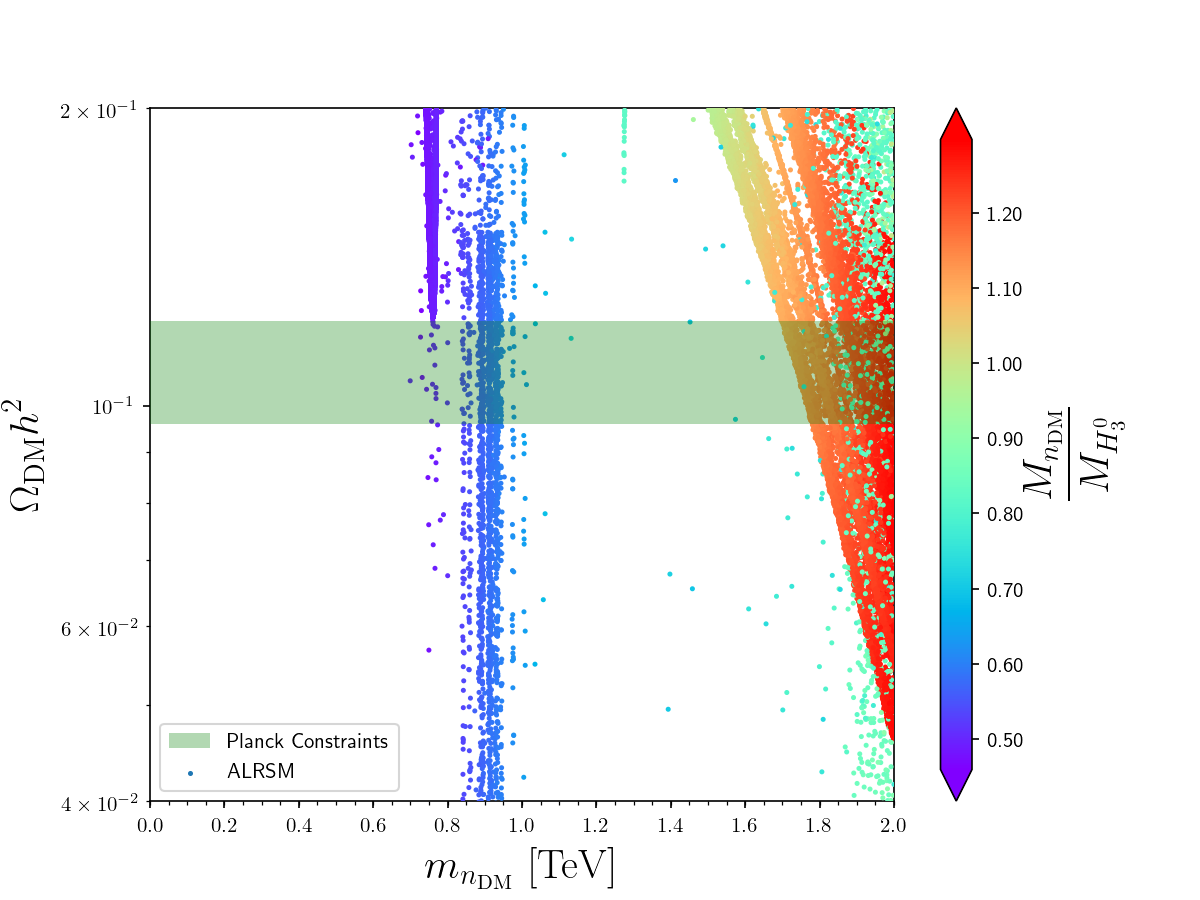}
  \includegraphics[width=.49\columnwidth]{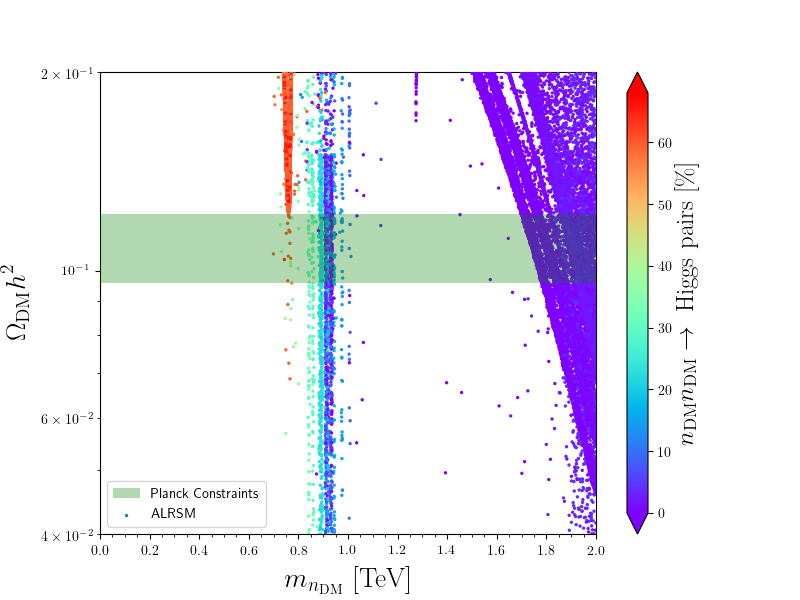}
  \includegraphics[width=.49\columnwidth]{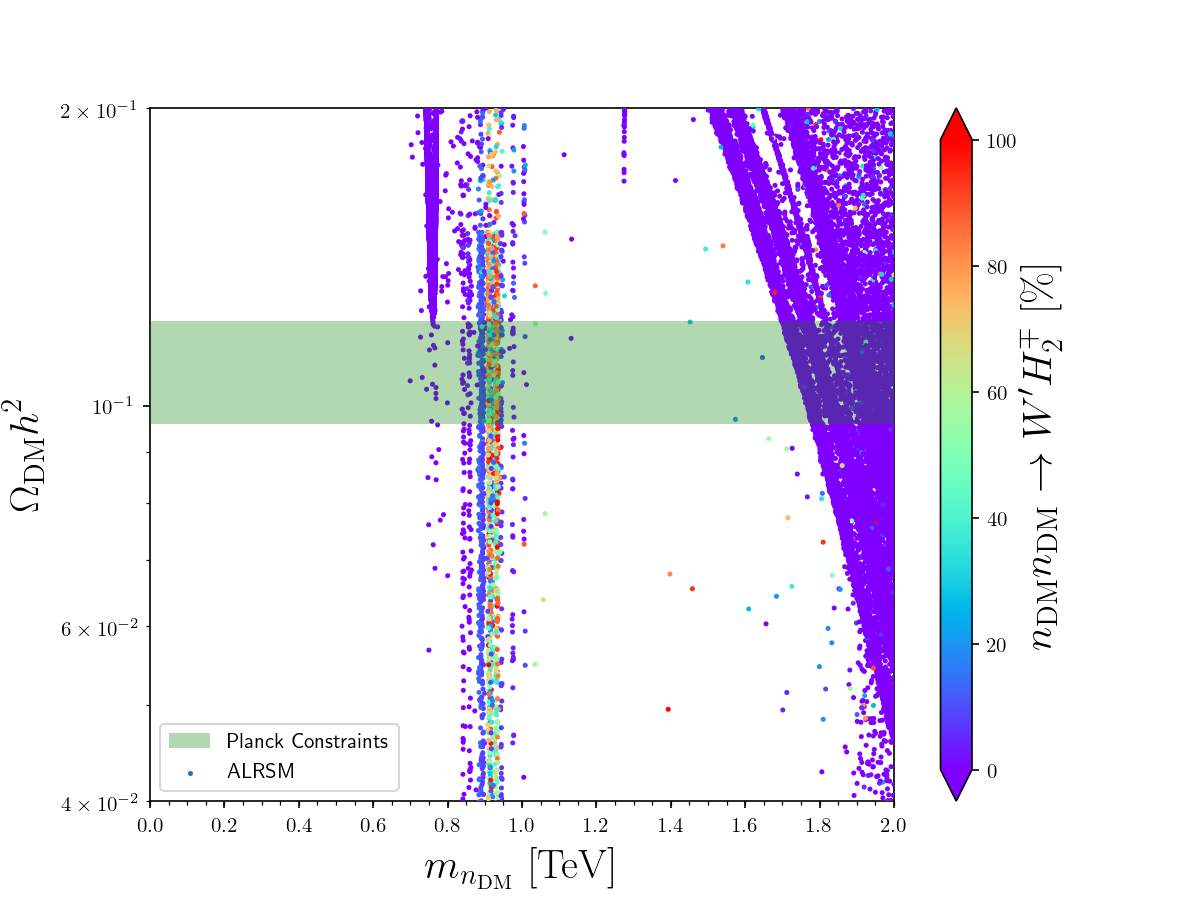}
  \includegraphics[width=.49\columnwidth]{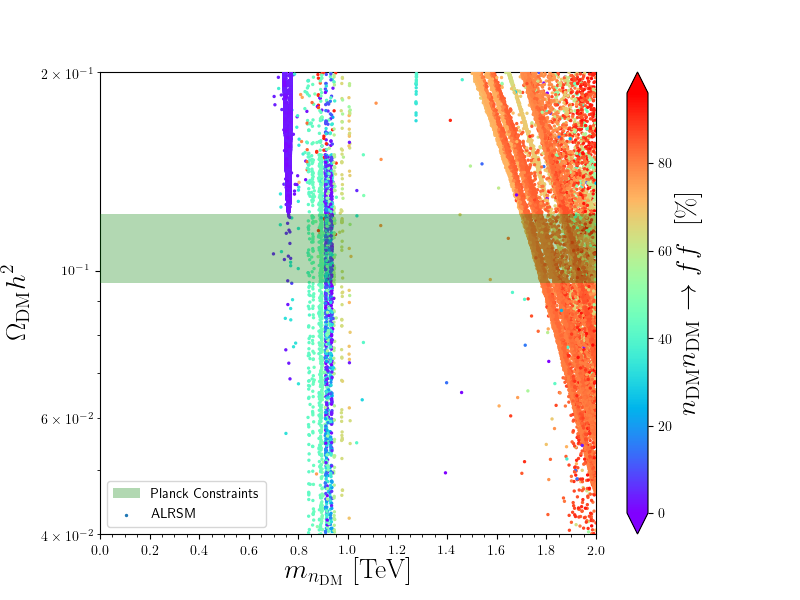}
  \caption{Relic density predictions for all ALRSM scenarios satisfying the
    Higgs constraints imposed during our scan and compatible with LHC $Z'$
    bounds, and its dependence on the mass of the lightest scotino. In each
    panel of the figure, we depict a specific property of all those scenarios.
    In the upper left panel, we represent by a colour code the mass of the
    $H_1^0$, $A_1^0$ and $H_2^\pm$ Higgs states, whilst in the upper right
    panel, we focus on the one of the $H_2^0$ and $A_2^0$ Higgs bosons. The mass
    of the scalar Higgs boson $H_3^0$ is presented relatively to the scotino
    mass in the central left panel, and the fractions of the DM annihilation
    cross section associated with annihilations in Higgs bosons, ${W'}^\pm
    H_2^\mp$ systems and fermions pairs are given in the central right, lower
    left and lower right panels respectively.}
  \label{fig:ALRM_DM}
\end{figure}

In this section, we investigate the constraints on the model arising from
imposing the lightest scotino as a viable DM candidate with properties
compatible with current cosmological data. First, we require that the predicted
relic density agrees within 20\% (to conservatively allow for uncertainties on
the predictions) with the recent Planck results, $\Omega_{\rm DM} h^2 =
0.12$~\cite{Ade:2013zuv}. We calculate, for all points returned by our scanning
procedure that are in addition compatible with the LHC $Z'$-boson bounds (see
section~\ref{sec:gaugebosons}), the associated DM relic density. We present our
results in figure~\ref{fig:ALRM_DM}. In all the subfigures, the
relic density is given as a function of the mass of the lightest scotino that we
denote by $m_{n_{\rm DM}}$. Two classes of solutions emerge from the results. In
a first set of allowed masses, the lightest scotino is quite light, with a mass lying
in the $[700, 1050]$~GeV window. The relic density as observed by the Planck
collaboration can however also be accommodated when the spectrum is heavier, \ie\
with a lightest scotino featuring $m_{n_{\rm DM}} \in [1.7, 2]$~TeV. This last
case is naturally less appealing from a collider search point of view. For this reason,
we did not increase the scanned scotino mass range (see section~\ref{sec:scan}),
although potentially viable scenarios could be obtained for even heavier
scotinos, and we mostly ignore this regime in the following discussion. In this
case, the right value obtained for the relic density prediction stems from
enhanced annihilations into fermions through $Z'$-boson $s$-channel
exchanges (see the lower right panel of the figure).

In the different panels of figure~\ref{fig:ALRM_DM}, we analyse the properties of
those ALRSM scenarios for which a relic density compatible with Planck data has
been found. A first remarkable feature is that when the DM scotino state is
light (\ie\ when $m_{n_{\rm DM}} \in [700, 1050]$~GeV), several Higgs bosons are
also light (upper left panel of the figure). The degenerate $H_1^0$ and $A_1^0$
neutral states, as well as the charged $H_2^\pm$ boson, hence have masses of
100--200~GeV. The heavier  the lightest scotino, the lighter 
these scalar and pseudoscalar bosons turn out to be. More precisely, for a scotino mass of
about 750~GeV, the (pseudo)scalar masses are  about 200~GeV, whilst for a
scotino mass of 800--1000~GeV, they turn out to be about
100~GeV. Moreover, the second scalar states $H_2^0$ and $A_2^0$ are only
slightly heavier
(upper right panel of figure~\ref{fig:ALRM_DM}), with masses found to lie
around 400~GeV. As a consequence of the presence of all those light states,
scotino annihilations into pairs of Higgs bosons contribute significantly to the
total annihilation cross section, as illustrated in the central right panel
of figure~\ref{fig:ALRM_DM}. This figure shows that on the contrary to any other
regime probed in our scan, channels where DM annihilates into Higgs bosons
contribute about 30--65\% to the total relic density when $m_{n_{\rm DM}} \in
[700, 1050]$~GeV. Such an enhancement (by comparison with heavier DM scenarios
where those channels are usually negligible) arises from the heaviest scalar
state $H_3^0$ that can mediate several DM annihilation modes. This scalar
boson is found to have a mass roughly equal to twice the DM mass $M_{H_3^0}
\approx 2 m_{n_{\rm DM}}$ (see the central left panel of
figure~\ref{fig:ALRM_DM}). There hence exists a new funnel allowing for
efficient DM annihilations into Higgs bosons, preventing DM from being
over-abundant. In addition,
the $H_3^0$ funnel also mediates annihilations into $W'{}^\mp H_2^\pm$ systems,
that turn to be dominant for a DM mass of about 900~GeV (lower left panel of
figure~\ref{fig:ALRM_DM}).

\begin{figure}
  \centering
  \includegraphics[width=.70\columnwidth]{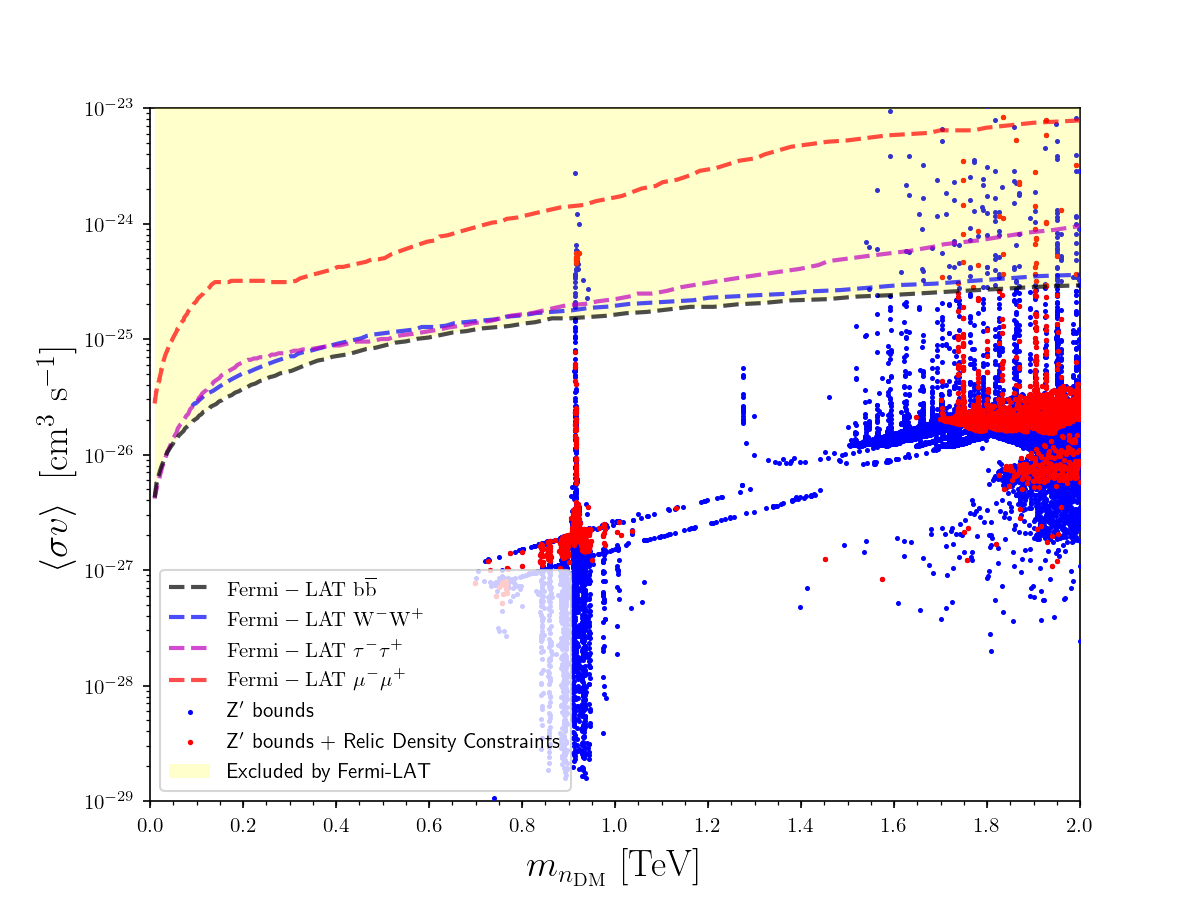}
  \caption{Predictions for the total DM annihilation cross section as a function
    of the mass of the lightest scotino. We show all points returned by the
    scan and that are compatible with LHC $Z'$ bounds. Scenarios for which the
    predictions for the relic density agree with Planck data are shown in red,
    whilst scenarios for which DM is over-abundant or under-abundant are shown
    in blue. We superimpose to our predictions constraints from
    Fermi-LAT~\cite{Ahnen:2016qkx}, the yellow area being excluded.}
  \label{fig:ID}
\end{figure}

Whilst we have demonstrated that the lightest scotino could be a viable DM
candidate from the point of view of the relic density, it is important to verify
that dark matter indirect and direct detection bounds are at the same time
satisfied. In figure~\ref{fig:ID}, we present the value of the total DM
annihilation cross section at zero velocity as a function of the scotino mass
for all scanned scenarios satisfying the $Z'$-boson LHC limits.
Configurations for which the relic density is found in agreement with Planck
data are shown in red, whilst any other setup returned by the scan is shown in
blue. In our predictions, we have moreover rescaled the DM annihilation cross
section to its present-day density. We compare our predictions to the latest
bounds derived from the Fermi satellite mission data~\cite{Ahnen:2016qkx}.
We depict, as a yellow area, the parameter space region that is found out to be
excluded. Most scanned scenarios naturally feature an annihilation
cross section that is  1 or 2 orders of magnitude too small to leave any potentially
visible signals in Fermi-LAT data, with a few exceptions where the
annihilation cross section at present time is enhanced. In general, such an
enhancement simultaneously leads to a reduction of the relic density so that
Planck data is at the same time accommodated. Equivalently, a significant
fraction of the scenarios that are excluded by indirect detection bounds turn
out to feature a relic density agreeing with cosmological data (the red points
lying within the yellow contour). Fortunately, most  potentially viable parameter regions 
from the relic density standpoint are  unaffected by current indirect
detection limits and will potentially stay so for some time by virtue of their
correspondingly small annihilation cross sections.

\begin{figure}
  \centering
  \includegraphics[width=.49\columnwidth]{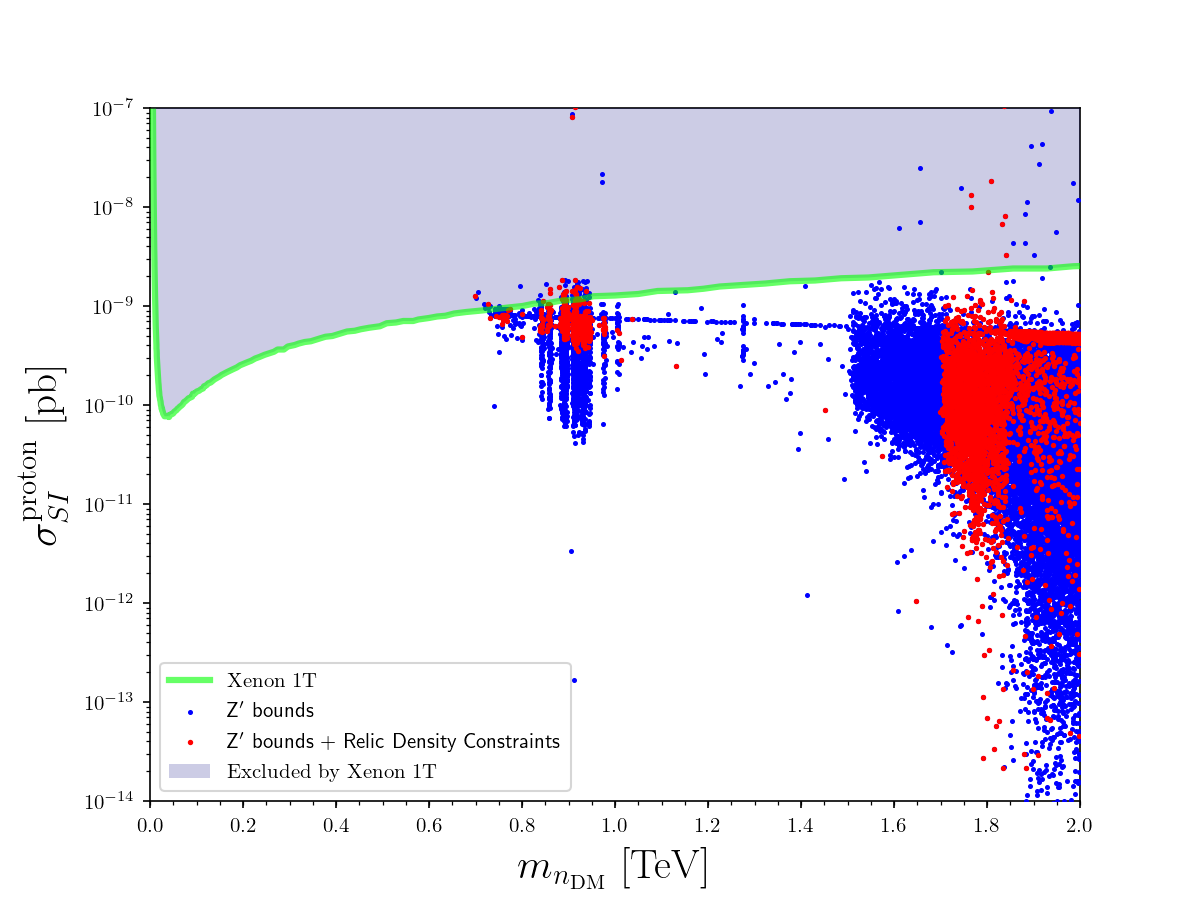}
  \includegraphics[width=.49\columnwidth]{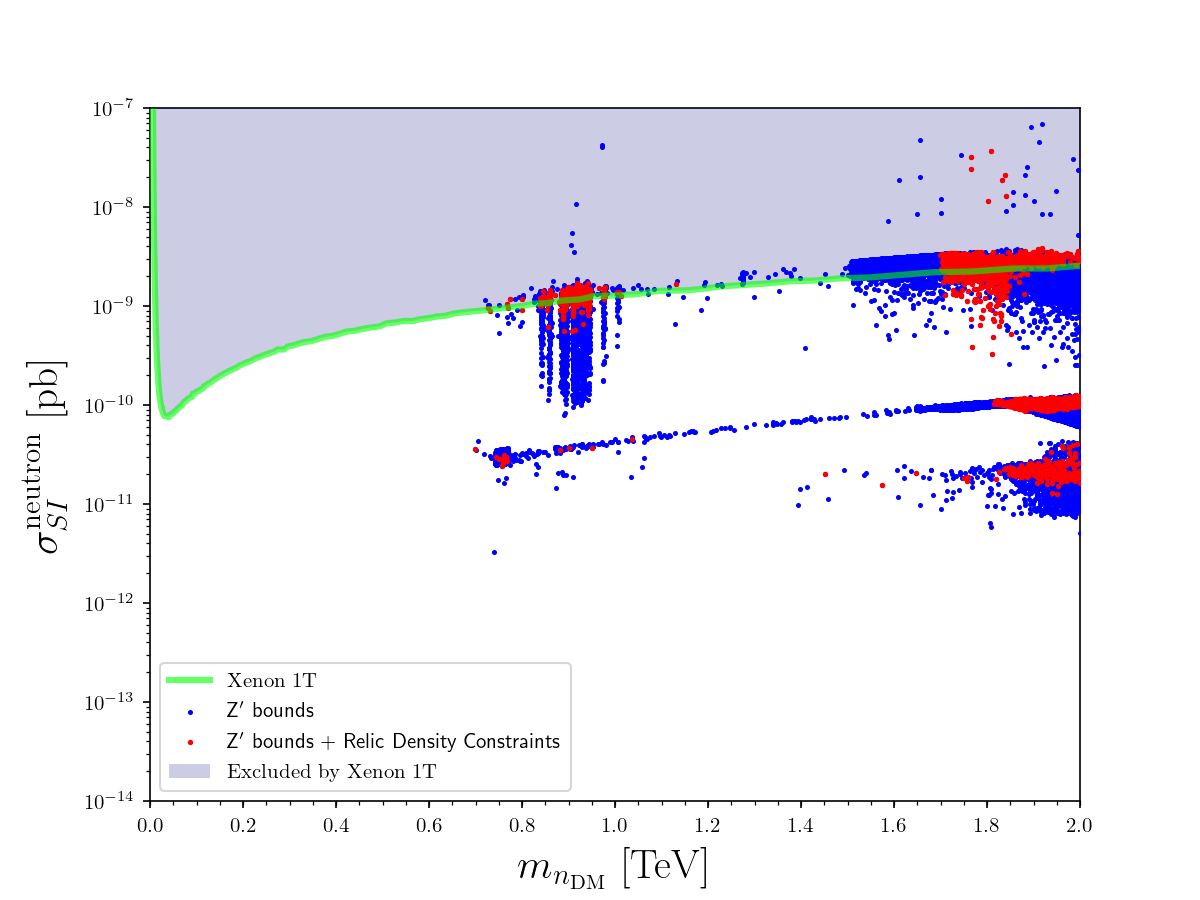}
  \caption{DM-proton (left) and DM-neutron (right) spin-independent scattering
    cross section as a function of the mass of the lightest scotino $m_{n_{\rm
    DM}}$.  Red points represent the scenarios featuring a relic density
    consistent with Planck data, and  blue point any other scenario returned
    by the scan. We restrict the results to scenarios satisfying the LHC $Z'$
    bounds.}
  \label{fig:DD}
\end{figure}

In figure~\ref{fig:DD}, we focus on DM direct detection bounds and represent the
DM-proton (left panel) and DM-neutron (right panel) spin-independent scattering
cross section $\sigma_{\rm SI}^{\rm proton}$ and $\sigma_{\rm SI}^{\rm neutron}$
as a function of the of the mass of the lightest scotino. Once
again, our results are normalised to the present-day relic density and points
compatible (incompatible) with Plank data are shown in red (blue). Our
predictions are then compared with the results of the Xenon~1T experiment~\cite{
Aprile:2018dbl}. In the ALRSM, neutron-scotino scattering cross sections are
naturally larger than proton-scotino scattering ones by virtue of the differences
between the $Z$ and $Z'$ couplings to the up-type and down-type quarks, so that
stronger constraints arise from the former process. Moreover, the distribution
of points in three clusters, as visible in the right panel of
figure~\ref{fig:DD}, stem from two features. First, these clusters are 
associated with different $Z'$ mass ranges, lighter $Z'$-bosons being associated
with smaller neutron-DM scattering rates. Second, down-type quarks play a
special role in the ALRSM as they do not couple to the $Z'$-boson. This impacts
the DM-neutron scattering cross section (consequently due to the larger down-quark
content of the neutron) whilst leading to a more `continuous' behaviour for the
DM-proton scattering cross section. A large fraction of all scenarios
accommodating the correct relic density are consequently excluded by the Xenon~1T
limits on the neutron-DM scattering cross section. Few options featuring a
scotino mass in the 700--1050~GeV range survive, made possible by a
suppression of the $Z'$-boson exchange diagrams due to a larger  $Z'$ boson mass
 in those scenarios.

In conclusion, we were able to obtain scenarios satisfying DM relic density and
direct and indirect detection constraints. The existence of those scenarios is
however pretty constrained, in particular due to direct detection bounds that
put severe requirements on the model spectrum, rendering it very predictable. In the surviving scenarios, the lightest
scotino (\ie\ our DM candidate) has a mass in the 750--1000 GeV window and a set
of non-SM-like Higgs bosons are light. In particular, the lightest $H_1^0$ and
$A_1^0$ bosons, as well as the $H^\pm_2$ boson, have masses in the 100--200~GeV
window. Moreover, the next scalar state $H_2^0$ and pseudoscalar state $A_2^0$
are only mildly heavier, with masses in general around 400~GeV. The heaviest
scalar $H_3^0$ is in contrast much heavier, with a mass roughly equal to twice
the lightest scotino mass. As a consequence of the presence of the funnel
topology, the DM annihilation cross section is predicted to be in the right range of
values to accommodate Planck data. A small fraction of scenarios are moreover
compatible with DM direct and indirect detection bounds. Another general feature
is that those scenarios feature a potentially  light $W'$ boson, with a mass
lying in the 1--2~TeV range,  not excluded by the results of the LHC.

\section{Scotino DM signal at colliders}
\label{sec:collider}
\begin{table}[t]
  \renewcommand{\arraystretch}{1.3}\setlength\tabcolsep{6pt}
  \begin{center}
    \begin{tabular}{c|c c c c c c}
       &$\tan\beta$& $\gR$ & $v'$ [GeV] & $\lambda_3$ & $\kappa$ [GeV]
         & $\alpha_1 = \alpha_2 = \alpha_3$\\
       \hline\hline
      {\bf BM I}   &4.58 & 0.374 & 7799 & 0.0196 & -31.08 & 0.0144 \\
      {\bf BM II}  &1.78 & 0.370 & 6963 & 0.0237 & -2.43  &  0.110 \\
      {\bf BM III} &4.55 & 0.374 & 7799 & 0.0196 & -30.38 & 0.0144 \\
    \end{tabular}\\[.2cm]
    \begin{tabular}{c|c c c c c c c}
       [GeV] & $M_{H_1^0}$   & $M_{H_2^0}$  & $M_{H_3^0}$
       & $M_{A_1^0}$   & $M_{A_2^0}$
       & $M_{H_1^\pm}$ & $M_{H_2^\pm}$ \\
       \hline\hline
      {\bf BM I}   & 193 & 907 & 1546 & 193 & 907 & 907 & 194\\
      {\bf BM II}  & 82  & 213 & 1578 & 82  & 167 & 167 & 82\\
      {\bf BM III} & 192 & 894 & 1546 & 192 & 894 & 894 & 192\\
    \end{tabular}\\[.2cm]
    \begin{tabular}{c|c c c c c c c c}
       [GeV] & $M_{Z'}$  & $M_{W'}$
       & $M_{n_1}$ & $M_{n_2}$ & $M_{n_3}$
       & $M_{d'}$  & $M_{s'}$  & $M_{b'}$ \\
       \hline\hline
      {\bf BM I}   & 4992 & 1460 & 756 & 971  & 1202 & 1500 & 1800 & 2000\\
      {\bf BM II}  & 5113 & 1288 & 909 & 1134 & 1223 & 1400 & 1822 & 2200\\
      {\bf BM III} & 4992 & 1460 & 902 & 1023 & 1312 & 1500 & 1936 & 2821\\
    \end{tabular}
    \caption{Values of the free ALRSM parameters defining our three
      benchmark scenarios {\bf BM I}, {\bf BM II} and {\bf BM III} (upper panel)
      and resulting mass spectrum (middle and lower panels). All masses are given in GeV.}
  \label{tab:BenchmarkFree}
  \end{center}
\end{table}

In this section we explore the implications at the LHC of the cosmology-favoured
scenarios that have emerged from our dark matter analysis. We choose three
benchmark scenarios consistent with the constraints previously studied and
provide their definition in terms of the model free parameters in the upper
panel of table~\ref{tab:BenchmarkFree}. As detailed in section~\ref{sec:scan},
the scalar potential parameter $\lambda_2=0$ for all scenarios. Moreover, the
small $\lambda_3$ value, together with the equality of all $\alpha_i$ parameters
and the moderate $\kappa$ value, implies that the $A_1^0$, $H_1^0$ and $H_2^\pm$
Higgs bosons are quite light (as derived from the relations presented in
appendix~\ref{app:higgs}). We have also chosen scenarios with a small $\gR$
value close to the theoretically allowed limit, which guarantees a light $W'$-boson (see section~\ref{sec:gaugebosons}) and induces $v'\approx\vR\sim7-8$~TeV.
The breaking of the $SU(2)_{R'}\times U(1)_{B-L}$ symmetry at such a scale
naturally leads to a $Z'$-boson mass of about 5~TeV for all benchmark scenarios
and a $W'$-boson mass of about 1.5~TeV. This is more precisely shown in the lower
and middle panels of table~\ref{tab:BenchmarkFree} in which we present the
masses of all new physics fields. In the selection of our benchmark points, we
impose  the lightest scotino to have a mass in the [700--1050]~GeV mass window,
the {\bf BM~I} scenario focusing on a lighter DM option ($m_{n_{\rm DM}} \approx
750$~GeV) and the two other scenarios on a heavier setup ($m_{n_{\rm DM}}\approx
900$~GeV). As discussed in section~\ref{sec:dm}, many Higgs states are quite
light, with masses of about 200~GeV ({\bf BM~I} and {\bf BM~III} scenarios) or
100~GeV ({\bf BM~II} scenario). In addition, our benchmark points choice is
LHC-driven, so that we target spectra in which the exotic down-type quarks are
heavier than the $W'$-boson so that a typical model signature could consist of
$W'$-boson pairs produced in association with jets through the $p p \to d'd' \to
W' j W' j$ process, for instance.

An interesting feature  of the model concerns the lightest
charged Higgs boson $H_2^\pm$, that, from the LHC perspective, is long-lived, so
that previous studies~\cite{Ashry:2013loa} are inapplicable.
As seen in table~\ref{tab:DecayModes}, the $H^{\pm}_2$ decay width is indeed of about
$2\times 10^{-18}$~GeV for the {\bf BM~I} and {\bf BM~III} scenarios, and of
$2\times 10^{-20}$~GeV for the {\bf BM~II} case, so that those scenarios could be
probed by searches for heavy stable charged particles (HSCP), the
$H_2^\pm$ bosons being pair-produced via the Drell-Yan mechanism. The
corresponding cross sections are given in table~\ref{tab:DecayModes}, for
proton-proton collisions at centre-of-mass energies of 7, 8 and 13~TeV and
for electron-positron collisions at a centre-of-mass energy of 183~GeV.
As the $H_2^\pm$ boson is lighter in the {\bf BM~II} scenario than in the other
two scenarios, the associated predictions are larger in the {\bf BM~II} case.
For instance, for proton-proton collisions at 13~TeV, the total production rate
hence reaches about 414 fb, compared to about 18~fb for the {\bf BM~I} and
{\bf BM~III} cases.

The related searches in 13~TeV LHC collisions exclude signal cross sections
ranging from 10 to 100~fb, the exact limit value depending on the
model~\cite{CMS:2016ybj,Khachatryan:2016sfv,Aad:2020srt,Aaboud:2019trc,Aaboud:2017iio,Alimena:2019zri}. 
The cross sections associated with
{\bf BM~I} and {\bf BM~III} $H_2^\pm$-boson pair
production lie at the border of the stau exclusion limits, so that it is
possible that two those benchmark scenarios are excluded. However, a direct
transposition of the limits is not straightforward as a consequence of the
modeling of various detector effects, which renders any conclusive
statement complicated. Similar conclusions hold for 7 and 8~TeV LHC search
results~\cite{Chatrchyan:2013oca,Khachatryan:2011ts,Aad:2011hz}.
On the other hand, all those searches specifically target HSCP with masses
larger than 100 GeV, so that they are unsensitive to the {\bf BM~II}
scenario. For the latter, one must thus rely on LEP results, covering the
[45.9, 89.5]~GeV mass range~\cite{Ackerstaff:1998si}. Upper limits on typical
HSCP signal cross sections of 0.05--0.19~pb have been extracted from data, but
again for models different from
the one investigated in this work. Such a model dependence in the results once
again prevents us from reinterpreting the results in the ALRSM framework.
As HSCP search results may consist in a very general smoking gun on the model,
we strongly encourage the LHC experimental collaborations to provide
information allowing one to recast of their search precisely enough, as 
to be able to provide limits for the model considered in this work. In the
meantime, we focus on other probes for the model.

The heavier charged Higgs state $H_1^\pm$ could in principle be constrained by more standard searches for additional Higgs states,
such as the one of ref.~\cite{Sirunyan:2019hkq}. Those searches are however
always targeting a specific production mode and a given decay channel which are
not relevant in the cosmology-favoured ALRSM case. For example, 
the CMS \cite{Sirunyan:2019hkq} and ATLAS \cite{Aaboud:2016dig} collaborations have investigated the
LHC sensitivity to a charged Higgs boson decaying in the $H^\pm\to\tau^\pm\nu_\tau$
mode. In the heavy $H_1^\pm$ case (scenarios {\bf BM I} and {\bf BM III}), cross
sections of a few fbs are excluded whilst in the light case ({\bf BM~II}
scenario), the analysis targets charged Higgs boson production from the rare
decay of a top quark. For heavier charged Higgs bosons, analyses of charged
Higgs boson production and decay in a $tb$ final state or heavy Higgs boson production in association with a $tb$ pair or a $Wbb$ system have also been carried
on (see, {\it e.g.}, refs.~\cite{Sirunyan:2020hwv, Aaboud:2018cwk}).

We have compared, for all the experimentally relevant signatures, the
corresponding predictions (reported in table~\ref{tab:DecayModes}) in the
considered ALRSM scenarios with the most recent bounds. The cross sections
excluded at the 95\% confidence level have been found to be orders of magnitude larger
than our model predictions. Similarly, we have verified that the corresponding
mass ranges (for the heavy stable $H_2$ state) are not excluded
by LEP~\cite{Abbiendi:2013hk}.

The light neutral states $H_1^0$ and $A_1^0$ are also
long-lived, and can therefore leads to a missing-energy signatures (as they
cannot decay into lepton or quark pairs).  However, in the corresponding considered
spectrum, they can only be produced from rare decays of
exotic quarks, so that this gives rise to signatures
potentially worth investigating in order to discover or exclude the model.
In the following, we focus instead on more abundantly produced final states.

\begin{table}
  \renewcommand{\arraystretch}{1.3}\setlength\tabcolsep{6pt}
  \centering
  \begin{tabular}{l|c|c|c}
    Benchmarks & \textbf{BM I} & \textbf{BM II} & \textbf{BM III} \\
    \hline\hline
    $\Gamma{(H_1^\pm)}$[GeV]  & 3.07 & $1.9\times 10^{-3}$ & 3.07 \\
    $\sigma(pp \to H_1^\pm)$ @ 13 TeV [pb]& $ 6.503 \times 10^{-5}$ & $0.04352$ & $6.901 \times 10^{-5}$ \\
    $\sigma(pp \to H_1^\pm W^\mp b \bar{b})$ @ 13 TeV [pb]& $2.723 \times 10^{-3}$ & 2.44 & $2.919\times 10^{-3}$  \\ 
    $\sigma(pp \to H_1^\pm t \bar{b}+{\rm h.c.})$ @ 13 TeV [pb] & $2.664\times 10^{-3}$   & 2.374 & $2.859\times 10^{-3}$ \\
    \hline
    $\Gamma{(H_2^\pm)}$[GeV]  & $1.93\times 10^{-18}$ & $2.62\times 10^{-20}$ & $1.85\times 10^{-18}$ \\
    $\sigma(pp \to H_2^\pm H_2^\mp)$ @ 7 TeV [fb]   & 5.412 & 163.3 & 5.588 \\
    $\sigma(pp \to H_2^\pm H_2^\mp)$ @ 8 TeV [fb]   & 7.153 & 199.8 & 7.392  \\ 
    $\sigma(pp \to H_2^\pm H_2^\mp)$ @ 13 TeV [fb]  & $ 18.18$ & $414.7$ & $ 18.71$ \\ 
    $\sigma(ee \to H_2^\pm H_2^\mp)$ @ 183 GeV [fb] & - & 161.1 & -  \\                               
    \hline
     $BR (H_1^\pm \to t \bar{b}$)   & 99.6 \% & $-$ & 99.6 \% \\
     $BR (H_1^\pm \to W b \bar{b}$) & $-$ & 80.5 \% & $-$ \\
     $BR (H_1^\pm \to c \bar{s}$)   & $-$ & 8.9 \% & $-$ \\
     $BR (H_1^\pm \to \tau  \nu$)    & $-$ & 4.83 \% & $-$\\
     $BR( H_1^\pm \to c \bar{b}$)   & $-$ & 2.1 \% & $-$ \\
  \end{tabular}
  \caption{Properties of the light charged Higgs states for the {\bf BM I},
    {\bf BM II} and {\bf BM III} benchmark scenarios.}
  \label{tab:DecayModes}
\end{table}

\begin{table}
  \renewcommand{\arraystretch}{1.3}\setlength\tabcolsep{6pt}
  \begin{center}
    \begin{tabular}{c|c c c c}
       & $\Omega_{\rm DM} h^2$
       & $\sigma_{\rm SI}^{\rm proton}$ [pb]
       & $\sigma_{\rm SI}^{\rm neutron}$ [pb]
       & $\langle\sigma v\rangle$ [cm$^3$s$^{-1}$]\\
       \hline\hline
      {\bf BM I}   & 0.118 & $8.08 \times 10^{-10}$ & $2.88 \times 10^{-11}$
        & $7.81 \times 10^{-28}$\\
      {\bf BM II}  & 0.120 & $8.09 \times 10^{-10}$ & $8.37 \times 10^{-10}$
        & $3.29 \times 10^{-27}$\\
      {\bf BM III} & 0.119 & $7.72 \times 10^{-10}$ & $3.67 \times 10^{-11}$
        & $1.17 \times 10^{-27}$\\
    \end{tabular}
    \caption{Predictions, for the {\bf BM I}, {\bf BM II} and {\bf BM III}
      scenarios, of the observables discussed in our dark matter analysis of the
      previous section.}
  \label{tab:BenchmarkRelic}
  \end{center}
\end{table}

In table~\ref{tab:BenchmarkRelic}, we present, for each of
the considered benchmark scenarios, predictions for the dark matter features
studied in section~\ref{sec:dm}. Each scenario leads to predictions compatible
with the cosmological experimental bounds by virtue of a different dynamics. In
the first {\bf BM I} scenario, the DM annihilation cross section is dominated by
annihilations into Higgs-boson pairs ($\sim 60\%$) as well as into pairs of SM
gauge bosons ($\sim 35\%$), and fermions to a smaller extent. Such an
annihilation pattern is typical of light scotino DM setups, as illustrated
in the figure~\ref{fig:ALRM_DM}. In the {\bf BM II}
scenario, DM annihilates essentially in $W'{}^\mp H_2^\pm$ systems, whilst in
the {\bf BM III} scenario, it dominantly annihilates into pairs of SM charged
leptons ($\sim 50\%$), quarks ($\sim 30\%$) and neutrinos ($\sim 15\%$). The {\bf
BM II} and {\bf BM III} scenarios hence illustrate the two classes of viable
scenarios emerging from more moderately heavy scotino dark matter
($m_{n_{\rm DM}} \in [800, 1000]$~GeV).

\begin{table}
  \renewcommand{\arraystretch}{1.3}\setlength\tabcolsep{6pt}
  \begin{center}
    \begin{tabular}{c|c c c c}
       & $\sigma (pp \to Z')$    [fb] & $\sigma (pp \to W' W')$ [fb]
       & $\sigma (pp \to W' d')$ [fb] & $\sigma (pp \to d' d')$ [fb]\\
       \hline\hline
      {\bf BM I}   & 0.821 & 0.0458 & 0.574 & 1.65\\
      {\bf BM II}  & 0.871 & 0.0672 & 1.080 & 2.72\\
      {\bf BM III} & 0.810 & 0.0465 & 0.564 & 1.61\\
    \end{tabular}\\[.2cm]
    \begin{tabular}{c|c |c c c}
       & BR($Z'\to\ell\ell$) & BR($W'\to e\ n_{\rm DM}$)
       & BR($W'\to \mu\ n_{\rm DM}$) & BR($W'\to \tau\ n_{\rm DM}$)\\
       \hline\hline
      {\bf BM I}   & 0.166 & 0.203 & 0.054 & 0.020\\
      {\bf BM II}  & 0.167 & 0.158 & 0.056 & 0.016\\
      {\bf BM III} & 0.171 & 0.178 & 0.063 & 0.018\\
    \end{tabular}\\[.2cm]
    \begin{tabular}{c| c c c c}
       & BR($d'\to W'\ u$) & BR($d'\to W'\ c$)
       & BR($d'\to H_2^\pm\ u$) & BR($d'\to H_2^\pm\ t$)\\
       \hline\hline
      {\bf BM I}   & 0.764 & 0.041 & 0.089 & 0.047\\
      {\bf BM II}  & 0.919 & 0.049 & 0.014 & $\approx 0$\\
      {\bf BM III} & 0.764 & 0.041 & 0.089 & 0.048\\
    \end{tabular}
    \caption{Predictions, for the {\bf BM I}, {\bf BM II} and {\bf BM III}
      scenarios, of various quantities relevant for the associated LHC
      phenomenology at a centre-of-mass energy of 13~TeV. In our notation,
      $\ell$ equivalently denotes an electron or a muon.}
  \label{tab:lhc}
  \end{center}
\end{table}

In table~\ref{tab:lhc}, we show predictions relevant for the LHC phenomenology
at a centre-of-mass energy of 13~TeV for our three benchmark scenarios.
Production cross sections for various processes involving new physics states
are presented in the upper panel, whilst the middle and lower panels include the
dominant branching ratios of the extra gauge bosons and exotic down-type quarks.
We ignore monojet production via the associated production of a scotino pair
with a hard jet as this process occurs at a too small rate (${\cal O}(1)$~fb for
an optimistic 100~GeV requirement on the leading jet). Other new
physics processes generally occur at a larger rate, as shown in the
table. For all three scenarios, $Z'$-boson production is small enough relatively
to the LHC limits (by construction of our benchmarks). The rate is hence of
about 0.15~fb after accounting for the $Z'$-boson branching ratio into electron
and muon pairs, BR$(Z'\to\ell\ell) \sim 17\%$ for $\ell$ equivalently denoting
an electron or a muon. Consequently this makes the $Z'$ signal difficult to
observe, even with more luminosity. As the $W'$-boson only couples to SM up-type
quarks and exotic down-type quarks, it cannot be singly produced. We therefore
focus on other processes typical of the ALRSM that instead involve pairs of $W'$
bosons and exotic $d'$ quarks. The production of a pair of $W'$-bosons leads to
the production of multileptonic systems in association with missing transverse
energy carried away by scotinos, as illustrated by the branching ratio
information of the middle panel of table~\ref{tab:lhc}. The total $W'$-boson
branching ratio into leptons and scotinos BR$(W'\to\ell n_{\rm DM})$ 
reaches 20--30\% in all three scenarios, after including the subdominant
tau-lepton contribution. The resulting signal cross section (including the
branching ratio into a lepton-scotino pair) is then about 0.010~fb.
Such a rate is far beyond the reach of typical multileptons plus missing energy
searches at the LHC, as confirmed by reinterpreting~\cite{Dumont:2014tja,
Conte:2018vmg} and extrapolating~\cite{Araz:2019otb} the results of the CMS
search of ref.~\cite{Sirunyan:2017lae} targeting electroweak superpartner
production and decay in the leptons plus missing energy mode to 3~ab$^{-1}$ with
\ma\footnote{Details on the reimplementation of the CMS electroweak superpartner
search of ref.~\cite{Sirunyan:2017lae} in \ma\ can be found in refs.~\cite{
Chatterjee:2018gca,1676304}.}. This signal, featuring a production times decay rate observable in the 10~ab range
at the LHC (for a centre-of-mass energy of 13~TeV), could however become visible
at future colliders.

\begin{figure}
  \centering
  \includegraphics[width=0.41\columnwidth]{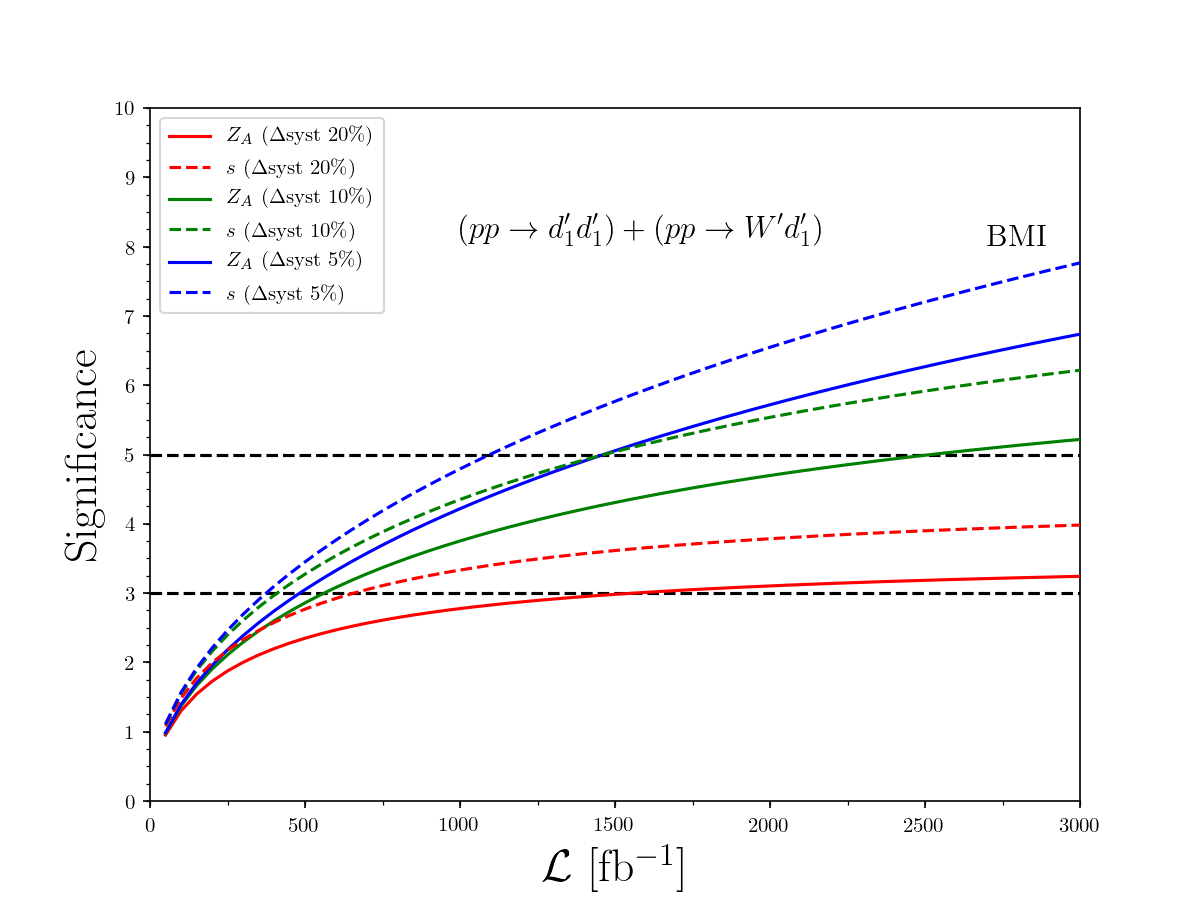}
  \includegraphics[width=0.41\columnwidth]{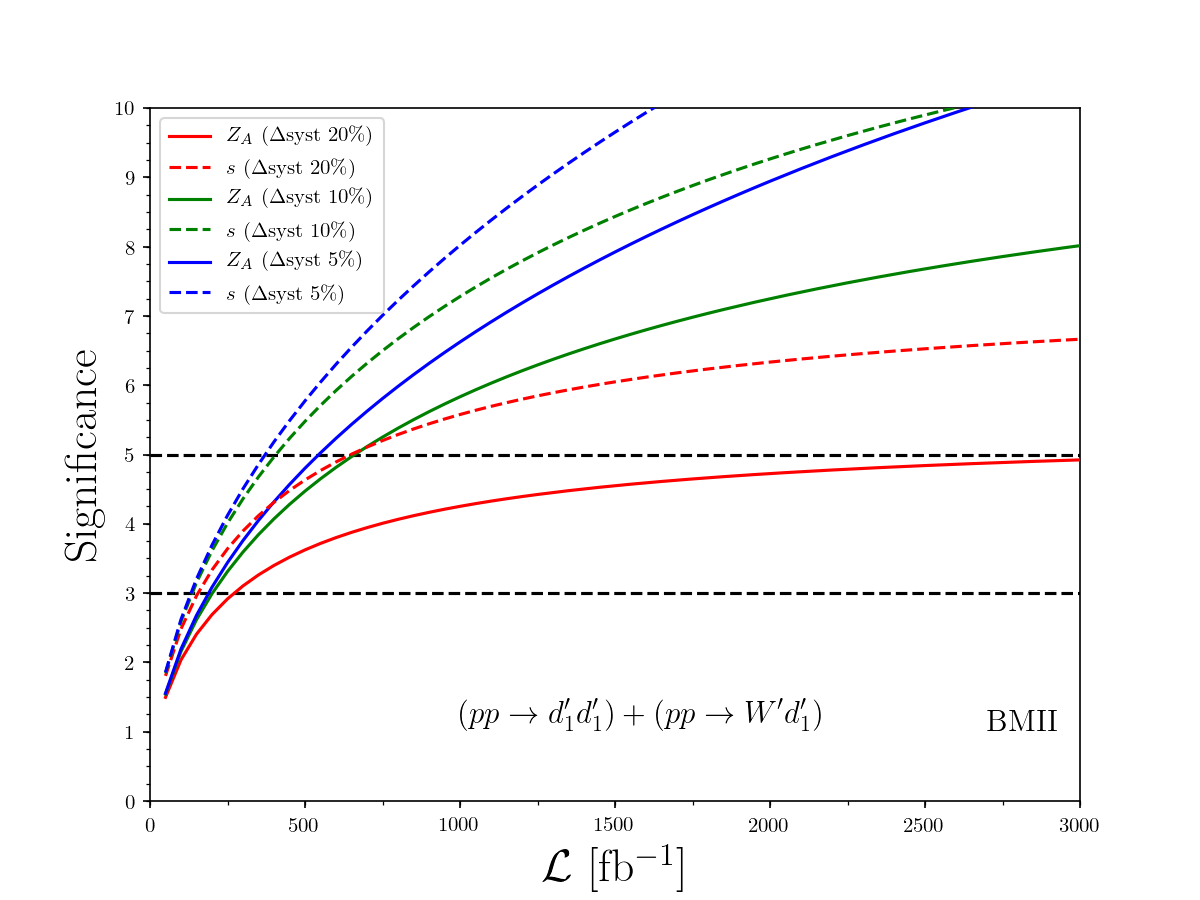}
  \includegraphics[width=0.41\columnwidth]{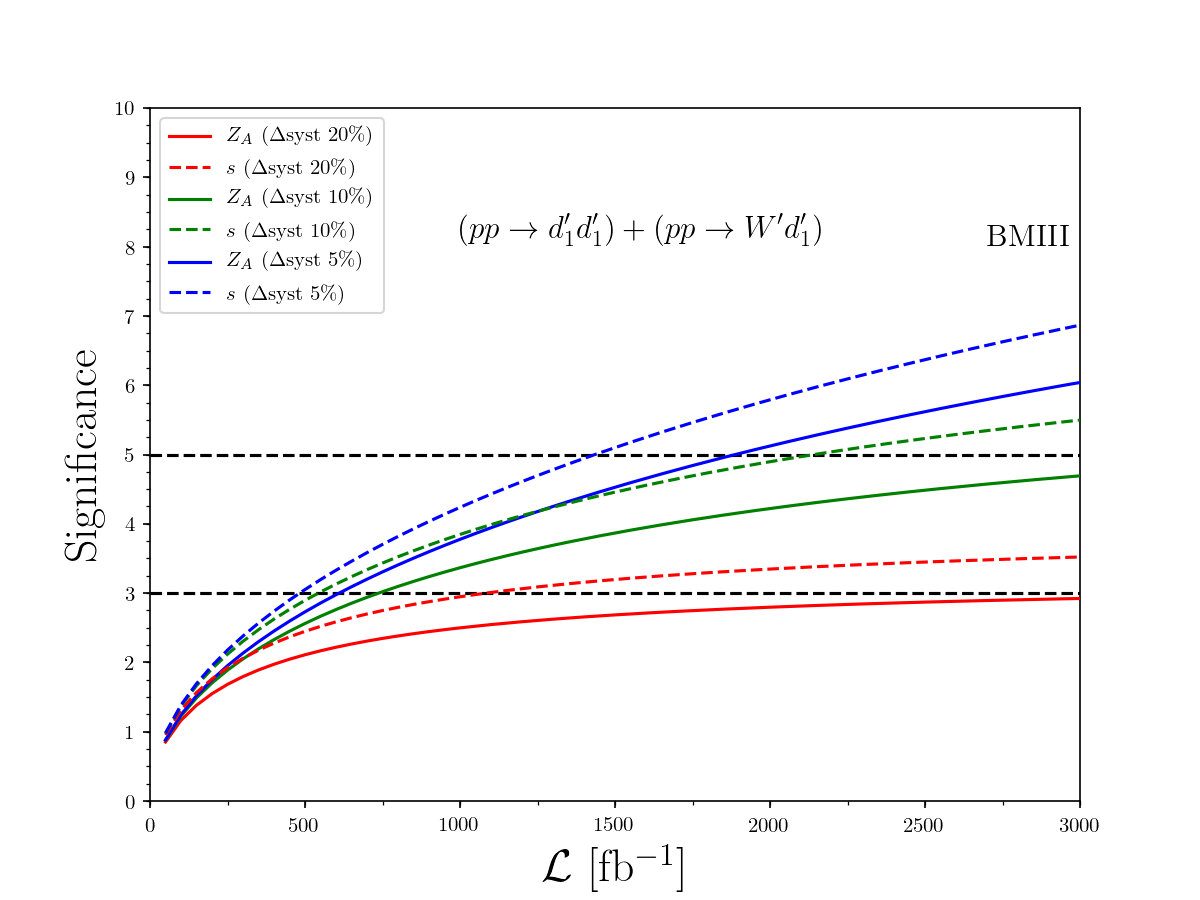}
  \caption{LHC sensitivity to a signature comprised of a dilepton, jets and
   missing energy in the context of the {\bf BM I} (upper left), {\bf BM II} (upper right) and
   {\bf BM III} (lower) scenarios. We present our results as a function of the
   luminosity and recast the CMS stop search of ref.~\cite{Sirunyan:2017leh},
  and plot the two significance measures of eq.~\eqref{eq:sign}.}
  \label{fig:sigIII}
\end{figure}

The upper panel of table~\ref{tab:lhc} also includes cross sections relevant for
$d'd'$ and $d'W'$ production. Such processes yield production cross sections in
the 1~fb range, which makes them potentially more appealing as a door to observing 
ALRSM at the LHC. Taking into account the large $d'\to W' j$ branching fraction,
a key signature of those processes is comprised of two leptons, jets and missing
transverse energy carried away by the scotinos emerging from the $W'$-boson
decays. This signature is also typically expected from supersymmetric squark
production and decay, so that the results of supersymmetry searches in the
opposite-sign dilepton, jets and missing energy mode could be reinterpreted to
constrain the ALRSM. We therefore recast the results of
the CMS stop search of Ref.~\cite{Sirunyan:2017leh} with \ma\footnote{Details on
the reimplementation of the CMS stop search of ref.~\cite{Sirunyan:2017leh} in
\ma\ can be found in refs.~\cite{Fuks:2018yku,1667773}.}, and extrapolate our
findings to 3~ab$^{-1}$. We present
our results in figure~\ref{fig:sigIII}. The LHC significance is evaluated
according to two measures, labelled by $s$ and $Z_A$, that are given by
\be\label{eq:sign}
  s    = \frac{S}{\sqrt{B + \sigma_B^2}} \quad\text{and}\quad
  Z_A  = \sqrt{2 \left[  (S+B) \ln \left[  \frac{(S+B)(S+\sigma_B^2)}{B^2+(S+B)\sigma_B^2 } \right] - \frac{B^2}{\sigma_B^2} \ln \left[ 1 + \frac{\sigma_B^2 S}{B(B+\sigma_B^2)} \right] \right]  } \ ,
\ee
where the number of selected signal and background events are denoted by $S$ and
$B\pm \sigma_{\rm B}$ respectively. The first method ($s$) is rather standard,
whereas the second one ($Z_A$) is more adapted to small numbers of background
events~\cite{Cowan:2010js}.  Moreover we consider a signal where both the
$W'd'$ and the $d'd'$ channels contribute. It turns out that while the LHC has currently  
very little sensitivity to the signal (\ie\ with 36~fb$^{-1}$), sensitivity levels
of about $3\sigma$ (for the {\bf BM~I} and {\bf BM~III} scenarios) to $5\sigma$
({\bf BM~II} scenario) could be reached at its high-luminosity operation phase
(\ie\ with 3000~fb$^{-1}$) with a conservative level of systematical
uncertainties of 20\%.
In the figure, we also show how a better understanding of the background
(corresponding to reduced uncertainties) could guarantee a discovery with
luminosities as low as about 750~fb$^{-1}$ (5\% of systematics) or
1500~fb$^{-1}$ (10\% of systematics) for the most optimistic {\bf BM~II}
scenario. For the two other more difficult to observe scenarios, the signal is suppressed so that
luminosities of about 1500-2000~fb$^{-1}$ should be necessary for a discovery with a level of
5\%  systematics.

\section{Summary and conclusions}
\label{sec:conclusion}

The Standard Model is plagued by several theoretical inconsistencies, while
being confirmed by experiments to a high degree of accuracy. Still, there are
at least two outstanding experimental facts which the SM does not explain:
neutrino masses and dark matter. The standard left-right symmetric model (LRSM)
naturally incorporates neutrino masses. However, without {\it ad hoc} additional
particles it does not include any viable dark matter candidate. We have
considered in this work an alternative realisation of the left-right symmetric model, the
so-called ALRSM, that can also be obtained from the breaking of an $E_6$ Grand
Unified setup. Such a class of models has the advantage to offer naturally
solutions for {\it both} neutrino masses and dark matter problems of the SM. 
Unlike in the LRSM, in ALRSM the $SU(2)_{R'}$ doublets of right-handed fermions
contain exotic states, namely down-type-like quarks $d'$ in the quark sector,
and neutrino-like scotinos $n$ in the lepton sector. The latter, being part of a
doublet, couples to the extra $W'$ and $Z'$ bosons. In this work, we have
shown that this property of the scotino is sufficient to promote it as a {\it
bona fide} dark matter candidate. Its gauge couplings indeed allow for a
sufficient increase in the DM annihilation cross section so that the relic
density, as measured by the Planck collaboration, can be accommodated.

Imposing various constraints on the model, such as requiring a cosmology
compatible with data (relic density, DM direct and indirect detection) and extra
gauge bosons not excluded by the LHC results, we have shown that scotino DM must
have a mass in a relatively narrow range of 750--1000~GeV (while ignoring heavier
options less appealing from the point of view of new physics at current collider
experiments). In addition, this restriction imposes strict mass bounds on several of the
Higgs bosons of the model. In particular, at least one scalar, one pseudoscalar
and one charged Higgs boson have to be light, in the 100--400~GeV mass regime.
Moreover, the $W^\prime$ gauge boson does not couple to pairs of ordinary
fermions so that its mass is mostly unconstrained, unlike the one of the $W_R$
boson of the usual LRSM. The only existing bounds arise indirectly, from limits
on the $Z^\prime$-boson mass derived from its non-observation in LHC data.
This however still allows the $W^\prime$ boson to be light, with a mass of
${\cal O}(1)$ TeV.
 The model also predicts additional light Higgs states. Given the
structure of the model, they however evade all present collider bounds. Of
these, a light charged Higgs boson is expected to be long lived, while neutral
states would manifest themselves as missing transverse energy at colliders.

We have devised three benchmark scenarios and studied the possibility of
observing those DM-favoured ALRSM realisations at the LHC. We have tested the
relevance of the ALRSM signatures arising from the $pp \to W' W'$, $W'd'$ and
$d'd'$ processes. For our choice of spectra, we have shown that the latter two
processes have similar cross sections, so that they could both provide an opportunity for the 
discovery of the ALRSM at the LHC. Out of the three benchmarks, the most
promising one can indeed yield a $5\sigma$ discovery within the future
high-luminosity run of the HL-LHC, the exactly luminosity  needed depending on
assumptions made on the systematic errors. The two other scenarios, associated
with smaller cross sections, are harder to probe but good prospects are foreseen
provided one gets a better control of the background. On the other
hand, HSCP searches could possibly consist in smoking guns on the model,
provided that future results are either directly interpreted in the ALRSM
framework or are released together with enough information for a proper
recasting.

In summary, the ALRSM analysed here has numerous attractive features once we impose that its
cosmological properties accommodate data: light Higgs bosons, a light charged
gauge boson, neutrino masses, and a viable dark matter candidate. The latter in
particular renders the spectrum well-defined. In addition, such ALRSM scenarios
emerge naturally from a grand unified $E_6$ theory, a promising
UV completion of the SM, and they offer the promise of being detectable at the
high-luminosity LHC.

\section*{Acknowledgments}
The authors are grateful to S.~Chakraborti and P.~Poulose and for enlightening
discussions in the early stages of this work, and to J.~Araz for useful comments
and suggestions. Parts of our numerical calculations have been performed using
the High Performance Computing (HPC) server managed by Calcul Qu{\'e}bec and
Compute Canada. The work of MF and \"{O}\"{O} has been partly supported by NSERC
through the grant number SAP105354, and the work of BF has been partly supported
by French state funds managed by the Agency Nationale de la Recherche (ANR), in
the context of the LABEX ILP (ANR-11-INDEX-0004-02, ANR-10-LAB-63). \"{O}\"{O}
thanks the University of Southampton, where part of
this work was completed, for their hospitality.

\appendix
\section{Diagonalisation of the scalar sector}
\label{app:higgs}
The scalar potential $V_{\rm H}$ of eq.~\eqref{eq:Hpot} is  bounded from below
if
\be
  \lambda_1 \geq 0\ ,\quad\lambda_2 \leq 0\ ,\quad\lambda_3\geq 0\ ,\quad
  \alpha_{12}\geq 0\ ,\quad\alpha_{13}\geq 0\quad\text{and}\quad
  \alpha_2-\alpha_3\geq 0\, ,
\label{eq:copos}\ee
where $\alpha_{ij} = \alpha_i  + \alpha_j$, and if one of the following
conditions is realised,
\be
 \Big[ \lambda_{12}\geq 0 \Big] \qquad\text{or}\qquad
 \Big[ \lambda_{12}\leq 0\ ,\quad \lambda_1+\lambda_2\geq 0
      \quad\text{and}\quad\lambda_1^2+4\lambda_2^2+8\lambda_1\lambda_2\leq 0
  \Big]\ ,
\label{eq:cstr2}\ee
with $\lambda_{12} = \lambda_1 + 2 \lambda_2$. Moreover, its minimisation allows
for the reduction of the number of degrees of freedom of the Higgs sector by
three,
\be
  \mu_1^2 = \alpha_{12} \big(\vL^2+\vR^2\big) + k^2 \lambda_1 +
    \frac{\kappa \vL \vR}{\sqrt{2} k} \ ,\qquad
  \mu_2^2 = \alpha_{12} k^2 + \lambda_3 \big(\vL^2+\vR^2\big) \ , \qquad
  \lambda_4 = \lambda_3 - \frac{\kappa k}{\sqrt{2} \vL \vR} \ .
\label{eq:minpot}\ee
Focusing first on the charged scalar sector, the squared mass matrix turns out
to be block diagonal. The $\phi^\pm_1$ and $\chi_L^\pm$ fields therefore mix
independently from the $\phi_2^\pm$ and $\chi_R^\pm$ fields, as shown by
eq.~\eqref{eq:ch_mix}. The corresponding $2\times 2$ blocks of the mass matrix
$({\cal M}^\pm_L)^2$ and $({\cal M}^\pm_R)^2$ are written, respectively, in
the $(\phi_2^\pm, \chi_L^\pm)$ and $(\phi_1^\pm, \chi_R^\pm)$ bases, as
\renewcommand{\arraystretch}{1.4}
\be\bsp
  ({\cal M}^\pm_{L,R})^2 =  \bpm
     -(\alpha_2-\alpha_3)v_{\scriptscriptstyle L,R}^2 -
         \frac{\kappa \vL \vR}{\sqrt{2}k}~~~~~ & 
     (\alpha_2-\alpha_3) k v_{\scriptscriptstyle L,R} +
         \frac{\kappa v_{\scriptscriptstyle R,L}}{\sqrt{2}}\\
     (\alpha_2-\alpha_3) k + \frac{\kappa v_{\scriptscriptstyle R,L}}{\sqrt{2}}&
     -(\alpha_2-\alpha_3)k^2 - \frac{\kappa k v_{\scriptscriptstyle R,L}}
       {\sqrt{2}v_{\scriptscriptstyle L,R}}\epm \ ,
\esp\ee
and are diagonalised by the rotations of eq.~\eqref{eq:ch_mix}. The
corresponding mass eigenvalues $M_{H_1^\pm}$ and $M_{H_2^\pm}$ are
\be
  M_{H_1^\pm} =  \frac{k^2+\vL^2}{2 k \vL} \Big[-2(\alpha_2-\alpha_3) k \vL -
    \sqrt{2} \kappa \vR\Big] \quad\text{and}\quad
  M_{H_2^\pm} =  \frac{k^2+\vR^2}{2 k \vR} \Big[-2(\alpha_2-\alpha_3) k \vR -
    \sqrt{2} \kappa \vL\Big] \ .
\ee
As $\alpha_2-\alpha_3\geq 0$ from eq.~\eqref{eq:copos}, forbidding tachyonic
fields yields $\kappa < 0$. This further implies $\lambda_4 \geq 0$ by virtue of
eq.~\eqref{eq:minpot}. As shown by eq.~\eqref{eq:nh_mix}, the pseudoscalar and
scalar components of the $\phi_1^0$ field do not mix and consist of the
physical $H_1^0$ and $A_1^0$ eigenstates. They are mass-degenerate, with masses
$M_{H_1^0}$ and $M_{A_1^0}$ reading
\be
  M^2_{H_1^0} = M^2_{A_1^0} = -(\alpha_2-\alpha_3) (\vL^2+\vR^2)
    - \frac{\kappa \vL \vR}{\sqrt{2} k} + 2 k^2 \lambda_2 \ .
\ee
The squared mass matrices $({\cal M}^0_\Re)^2$ and $({\cal M}^0_\Im)^2$ of the
three remaining scalar and pseudoscalar fields are respectively given, in the
$(\Re\{\phi^0_2\}, \Re\{\chi_L^0\}, \Re\{\chi_R^0\})$ and
$(\Im\{\phi^0_2\}, \Im\{\chi_L^0\}, \Im\{\chi_R^0\})$ bases, by
\renewcommand{\arraystretch}{1.3}
\setlength\arraycolsep{5pt}
\be\bsp
  ({\cal M}^0_\Re)^2 =&\ \bpm
    2k^2 \lambda_1 \!-\! \frac{\kappa \vL \vR}{\sqrt{2} k} & 2 \alpha_{12} k \vL
      \!+\! \frac{\kappa \vR}{\sqrt{2}} & 2 \alpha_{12} k \vR \!+\!
      \frac{\kappa \vL}{\sqrt{2}}\\
    2\alpha_{12} k \vL \!+\! \frac{\kappa \vR}{\sqrt{2}} & 2 \lambda_3 \vL^2
      \!-\!\frac{\kappa k\vR}{\sqrt{2} \vL} & 2 \lambda_3\vL\vR\!-\!
      \frac{\kappa k}{\sqrt{2}}\\
    2\alpha_{12} k \vR \!+\! \frac{\kappa \vL}{\sqrt{2}} & 2\lambda_3\vL\vR\!-\!
       \frac{\kappa k}{\sqrt{2}}  & 2 \lambda_3 \vR^2 \!-\!
       \frac{\kappa k \vL}{\sqrt{2} \vR}
  \epm \ , \\
  ({\cal M}^0_\Im)^2 =&\ \frac{\kappa}{\sqrt{2}} \bpm
    -\frac{\vL \vR}{k} & \vR & - \vL \\
    \vR & -\frac{k \vR}{\vL} & k\\
    -\vL & k & -\frac{k \vL}{\vR} \epm \ ,
\esp\ee
and are diagonalised by the two $U_{3\times 3}^{\rm H}$ and
$U_{3\times 3}^{\rm A}$ rotation matrices of eq.~\eqref{eq:nh_mix}. These are explicitly 
given by
\renewcommand{\arraystretch}{1.6}
\be\bsp
  U_{3\times 3}^{\rm A} = &\
  \frac{1}{\sqrt{2}}\bpm
    -\frac{k}{\sqrt{k^2+\vR^2}} & \frac{k \vR^2}{\sqrt{\big(k^2+\vR^2\big)
     \big(\vL^2 \vR^2+k^2 \vL^2 + \vR^2k^2\big)}} &
       \frac{\vR \vL}{\sqrt{\vL^2 \vR^2+k^2 \vL^2 + \vR^2k^2}}\\
     0 & \vL \frac{\sqrt{k^2+\vR^2}}{\sqrt{\vL^2 \vR^2+k^2 \vL^2 + \vR^2k^2}} &
      - \frac{k \vR}{\sqrt{\vL^2 \vR^2+k^2 \vL^2 + \vR^2k^2}}\\
     \frac{\vR}{\sqrt{\vR^2+k^2}} &
     \frac{k^2 \vR}{\sqrt{\big(k^2+\vR^2\big)
     \big(\vL^2 \vR^2+k^2 \vL^2 + \vR^2k^2\big)}} &
      \frac{k \vL}{\sqrt{\vL^2 \vR^2+k^2 \vL^2 + \vR^2k^2}}\epm\ ,\\
  U_{3\times 3}^{\rm H} = &\
    \frac{1}{\sqrt{2}}\bpm
    \frac{f_0}{\sqrt{{\cal D}_1}} &
    \frac{f_2(1+g_0^2) - f_0(1+g_0g_2)}{\sqrt{{\cal D}_1 {\cal D}_2}} &
    \frac{\xi (g_2-g_0)}{\sqrt{{\cal D}_2}}\\
    \frac{g_0}{\sqrt{{\cal D}_1}} &
    \frac{g_2(1+f_0^2) - g_0(1+f_0f_2)}{\sqrt{{\cal D}_1 {\cal D}_2}} &
    \frac{\xi (f_0-f_2)}{\sqrt{{\cal D}_2}}\\
    \frac{1}{\sqrt{{\cal D}_1}} &
    \frac{f_0^2 + g_0^2 - f_0f_2 -g_0g_2}{\sqrt{{\cal D}_1 {\cal D}_2}} &
    \frac{\xi (g_0f_2-g_2f_0)}{\sqrt{{\cal D}_2}}\epm \ ,
\esp\ee
and depend on various functions of the Higgs mass eigenvalues $M_{H^0_i}$,
\be\bsp
  f_i = &\ \frac{2 M_{H_i^0}^4 \vL \vR + M_{H_i^0}^2 (\vL^2 + \vR^2)
    (\sqrt{2} k \kappa - 4 \vL \vR \lambda_3) - 2 \sqrt{2} k (\vL^2 - \vR^2)^2
    \lambda_3 \kappa}{\vR \Big[M_{H_i^0}^2 (4 k \vL \vR \alpha_{12} +
    \sqrt{2} \vL^2 \kappa) + 2 \sqrt{2} (k^2 \alpha_{12} + \vL^2
    \lambda_3) (\vR^2 - \vL^2) \kappa\Big]} \ ,\\
  g_i = &\ \frac{\vL}{\vR}\ \frac{M_{H_i^0}^2 (4k \vL \vR \alpha_{12} + \sqrt{2}
    \vR^2 \kappa) + 2 \sqrt{2} (k^2 \alpha_{12} + \vR^2 \lambda_3) (\vL^2 -
    \vR^2) \kappa}{M_{H_i^0}^2 (4 k \vL \vR \alpha_{12} + \sqrt{2} \vL^2 \kappa)
    + 2 \sqrt{2} (k^2 \alpha_{12} + \vL^2 \lambda_3) (\vR^2 - \vL^2)\kappa}\ ,\\
  {\cal D}_1 =&\ 1+f_0^2+g_0^2 \ , \\
  {\cal D}_2 =&\ f_2^2 (1+g_0^2) + (g_0-g_2)^2 - 2 f_0 f_2 (1+g_0g_2) +
    f_0^2 (1+g_2^2) \ ,\\
  \xi = &\ {\rm sgn}\big[ g_0(f_2-f_3) + g_2(f_3-f_0)+g_3(f_0-f_2)\big] \ .
\esp\ee
In our conventions, we trade the $\lambda_1$ free parameter of the scalar
potential for the mass of the lightest Higgs state $H_0^0$ (that can then be set
freely and thus match the SM Higgs boson mass). $\lambda_1$ becomes thus a
dependent parameter,
\be\label{eq:lam1}
  \lambda_1 =  \frac{1}{2 k^3}\ \frac{
    \sqrt{2} k \vL \vR M_{H_0^0}^6 + \mathfrak{a}^{(4)} M_{H_0^0}^4
      - 2\mathfrak{a}^{(2)}M_{H_0^0}^2 - 4\alpha_{12}^2\kappa k^4(\vL^2-\vR^2)^2
  }{
    \sqrt{2} \vL \vR M_{H_0^0}^4
      + (\kappa k-2\sqrt{2}\lambda_3 \vL \vR) (\vL^2+\vR^2)M_{H_0^0}^2 
      - 2\kappa k \lambda_3 (\vL^2-\vR^2)^2
  }\ ,
\ee
and the remaining scalar masses then read
\be
  M^2_{A_2^0} =  - \frac{\kappa}{\sqrt{2} k \vL \vR}
    \Big[\vL^2 \vR^2 + k^2(\vL^2+\vR^2)\Big] \qquad\text{and} \qquad
 M_{H_{2,3}^0}^2 = \frac12 \bigg[ \mathfrak{a} \pm
   \sqrt{\mathfrak{a}^2 + 4(\mathfrak{b} + \mathfrak{a} M_{H_0^0}^2) }\bigg] \ .
\ee
with
\be\bsp
  \mathfrak{a}^{(4)} = &\ -2\sqrt{2} k \lambda_3 \vL \vR (\vL^2 + \vR^2) +
     \kappa \Big(\vL^2 \vR^2 + k^2 (\vL^2 + \vR^2)\Big) \ ,\\
  \mathfrak{a}^{(2)} = &\  2\sqrt{2} \alpha_{12}^2 k^3 \vL \vR (\vL^2 + \vR^2) +
     \kappa \Big(\lambda_3 \vL^2 \vR^2 (\vL^2 + \vR^2) + k^2 \big[4 \alpha_{12}
       \vL^2 \vR^2 + \lambda_3 (\vL^2 - \vR^2)^2\big]\Big) \  ,\\
  \mathfrak{a} =&\ \frac{1}{\sqrt{2} k \vL \vR} \bigg[
    \vL \vR \big(2 \sqrt{2} k^3 \lambda_1 - \kappa \vL \vR\big) +
    k \big(2 \sqrt{2} \lambda_3 \vL \vR -\kappa k\big)\big(\vL^2+\vR^2\big)
  \bigg]  -M_{H_0^0}^2\ , \\
  \mathfrak{b} =&\ \frac{1}{k \vL \vR} \bigg[
    \sqrt{2}\kappa k^2\Big(4 \alpha_{12} \vL^2 \vR^2 +
       \lambda_3 (\vL^2 \!-\! \vR^2)^2 \Big)\\ &\qquad\qquad+
    \Big(4 k^3 (\alpha_{12}^2 \!-\! \lambda_1 \lambda_3) \vL \vR +
       \sqrt{2}\kappa (k^4 \lambda_1 \!+\! \lambda_3 \vL^2 \vR^2)\Big)
       \Big(\vL^2 \!+\! \vR^2\Big) 
     \bigg] \ .
\esp\ee

\section{The fermion sector}
\label{app:ferms}
Fermion mass terms are generated from the Yukawa Lagrangian of
eq.~\eqref{eq:yuk} after the breaking of the $SU(2)_L\times SU(2)_{R'}\times
U(1)_{B-L}$ symmetry down to electromagnetism,
\be
  \lag_{\rm F}^{\rm mass} =
    -\frac{k}{\sqrt{2}}
      \Big[ \bar e_L {\bf \hat Y}^e e_R + \bar u_L {\bf \hat Y}^u u_R \Big]
    -\frac{\vL}{\sqrt{2}}
      \Big[ \bar d_L {\bf \hat Y}^d d_R + \bar \nu_L {\bf \hat Y}^\nu \nu_R\Big]
    -\frac{\vR}{\sqrt{2}}
      \Big[ \bar d'_R {\bf \hat Y}^{d'} d'_L + \bar n_R {\bf \hat Y}^n n_L \Big]
    + {\rm h.c.} 
\ee
The different mass matrices ${\bf \hat Y}$ can be diagonalised through 12
unitary rotations,
\be\renewcommand{\arraystretch}{1.}\bsp
  \frac{k}{\sqrt{2}}   {\bf\hat Y}^u
    \to \frac{k}{\sqrt{2}}   V_u {\bf Y}^u U_u^\dag
    =    \bpm M_u&0&0\\0&M_c&0\\0&0&M_t\epm \!,\
& \frac{\vL}{\sqrt{2}} {\bf\hat Y}^d
    \to \frac{\vL}{\sqrt{2}} V_d {\bf Y}^d U_d^\dag
    =    \bpm M_d&0&0\\0&M_s&0\\0&0&M_b\epm \!,\\
  \frac{\vL}{\sqrt{2}} {\bf\hat Y}^\nu
    \to \frac{\vL}{\sqrt{2}} V_\nu {\bf Y}^\nu U_\nu^\dag
    = \bpm M_{\nu_e}&0&0\\0&M_{\nu_\mu}&0\\0&0&M_{\nu_\tau}\epm \! ,\
& \frac{k}{\sqrt{2}}   {\bf\hat Y}^e
    \to \frac{k}{\sqrt{2}} V_e{\bf Y}^e U_e^\dag
    =   \bpm M_e&0&0\\0&M_\mu&0\\0&0&M_\tau\epm \! ,\\
  \frac{\vR}{\sqrt{2}}   {\bf\hat Y}^{d'}
    \to \frac{\vR}{\sqrt{2}} U_{d'}{\bf Y}^{d'} V_{d'}^\dag
    =   \bpm M_{d'}&0&0\\0&M_{s'}&0\\0&0&M_{b'}\epm \! ,\
& \frac{\vR}{\sqrt{2}}   {\bf\hat Y}^n
    \to \frac{\vR}{\sqrt{2}} U_n{\bf Y}^n V_n^\dag
    =   \bpm M_{n_e}&0&0\\0&M_{n_\mu}&0\\0&0&M_{n_\tau}\epm \! ,
\esp\ee
leading to diagonal and real ${\bf Y}$ matrices. These rotations
equivalently correspond to replacing the fermion gauge eigenbasis by the
physical one,
\be\bsp
 &  u_L \to V_u   u_L\ , \quad
    ~d_L \to V_d   d_L\ , \quad
  \nu_L \to V_\nu \nu_L\ , \quad
    e_L \to V_e   e_L\ , \quad
    ~d'_L\to V_{d'}d'_L\ , \quad
    n_L \to V_n   n_L\ , \\
 &  u_R \to U_u   u_R\ , \quad
    d_R \to U_d   d_R\ , \quad
  \!\nu_R \to U_\nu \nu_R\ , \quad
   \!e_R \to U_e   e_R\ , \quad
    d'_R\to U_{d'}d'_R\ , \quad
   \!n_R \to U_n   n_R\ .
\esp\ee
As in the SM, conventionally we  keep the left-handed up-type
quark and charged lepton bases unchanged and absorb the $V_u - V_d$ and $V_\nu
- V_e$ rotations in a redefinition of the down-type quark and neutrino states.
Similarly, the $U_u-U_{d'}$ and $U_n-U_e$ rotations are conveniently absorbed in
a redefinition of the $d'_R$ and $n_L$ bases, the right-handed up-type quark and
charged lepton bases being kept unchanged,
\be\bsp
  d_L \to V_u^\dag V_d d_L \equiv V_{\rm CKM} d_L \ , \quad&
  \nu_L \to V_e^\dag V_\nu \nu_L \equiv V_{\rm PMNS} d_L \ , \\
  d'_R \to U_u^\dag U_{d'} d'_R \equiv V_{\rm CKM'} d'_R\ , \quad&
  n_R \to U_e^\dag U_n n_R \equiv V_{\rm PMNS'} n_R \ .
\esp\ee
Omitting any potential Majorana phase, each of the four CKM/PMNS rotation
matrices can be defined by three mixing angles $\theta_{ij}$ and a Dirac phase
$\delta$.

\section{Technical details on our \fr\ implementation}
\label{app:fr}
\begin{table}
\centering
\renewcommand{\arraystretch}{1.4}
\setlength\tabcolsep{2pt}
\begin{tabular}{c c c c}
  Field & Spin & Name & PDG \\
  \hline\hline
  $Z'$                  & 1   & {\tt Zp}  & 32\\
  $W'^+$                & 1   & {\tt Wp}  & 34\\
  $n_i$ ($i=1,2,3$)     & 1/2 & {\tt nl}  & 6000012, 6000014, 6000016\\
  $d'_i$ ($i=1,2,3$)    & 1/2 & {\tt dqp} & 6000001, 6000003, 6000005\\
  $H^0_i$ ($i=0,1,2,3$) & 0   & {\tt h0}  & 25, 25, 45, 55\\
  $A^0_i$ ($i=1,2$)     & 0   & {\tt A0}  & 36, 46\\
  $H^+_i$ ($i=1,2$)     & 0   & {\tt Hp}  & 37, 47\\
\end{tabular}
\caption{Mass eigenstates that supplement the SM, together with their spin
  quantum number (second column), the name used in the \fr\ implementation
  (third column) and the associated PDG identifier (last column).}
\label{tab:fields}
\begin{tabular}{c c c c}
  Parameter& Name & LH block & LH counter\\
  \hline\hline
  $\tan\beta$ & {\tt tb}   & {\tt SMINPUTS} & 5\\
  $\gR$       & {\tt gR}   & {\tt SMINPUTS} & 6\\
  $v'$        & {\tt vevp} & {\tt SMINPUTS} & 7\\
  \hline
  $\lambda_2$ & {\tt lam2} & {\tt HPOTINPUTS} & 1\\
  $\lambda_3$ & {\tt lam3} & {\tt HPOTINPUTS} & 2\\
  $\alpha_1$  & {\tt alp1} & {\tt HPOTINPUTS} & 3\\
  $\alpha_2$  & {\tt alp2} & {\tt HPOTINPUTS} & 4\\
  $\alpha_3$  & {\tt alp3} & {\tt HPOTINPUTS} & 5\\
  $\kappa$    & {\tt kap}  & {\tt HPOTINPUTS} & 6\\
  \hline
  $a_{\rm H}^g $ & {\tt Ghgg} & {\tt EFFECTIVEHIGGS} & 1\\
  $a_{\rm H}^a $ & {\tt Ghaa} & {\tt EFFECTIVEHIGGS} & 2\\
\end{tabular}\hspace{.3cm}
\begin{tabular}{c c c c}
  Parameter& Name & LH block & LH counter\\
  \hline\hline
  $M_{\nu_e}$     & {\tt Mve}  & {\tt MASS} & 12\\
  $M_{\nu_\mu}$   & {\tt Mvm}  & {\tt MASS} & 14\\
  $M_{\nu_\tau}$  & {\tt Mvt}  & {\tt MASS} & 16\\
  $M_{n_e}$       & {\tt Mne}  & {\tt MASS} & 6000012\\
  $M_{n_\mu}$     & {\tt Mnm}  & {\tt MASS} & 6000014\\
  $M_{n_\tau}$    & {\tt Mnt}  & {\tt MASS} & 6000016\\
  $M_{d'}$        & {\tt MDP}  & {\tt MASS} & 6000001\\
  $M_{s'}$        & {\tt MSP}  & {\tt MASS} & 6000003\\
  $M_{b'}$        & {\tt MBP}  & {\tt MASS} & 6000005\\
\end{tabular}\\[.2cm]
\begin{tabular}{c c c c}
  Parameter& Name & LH block & LH counter\\
  \hline\hline
  $\lambda$       & {\tt CKMlam} & {\tt CKMBLOCK} & 1\\
  $A$             & {\tt CKMA}   & {\tt CKMBLOCK} & 2\\
  $\bar\rho$      & {\tt CKMrho} & {\tt CKMBLOCK} & 3\\
  $\bar\eta$      & {\tt CKMeta} & {\tt CKMBLOCK} & 4\\
  \hline
  $s_{12}^{\rm(CKM')}$ & {\tt CKMps12} & {\tt CKMBLOCK}& 11\\
  $s_{23}^{\rm(CKM')}$ & {\tt CKMps23} & {\tt CKMBLOCK}& 12\\
  $s_{13}^{\rm(CKM')}$ & {\tt CKMps13} & {\tt CKMBLOCK}& 13\\
  $\delta_{\rm CKM'}$  & {\tt CKMpdel} & {\tt CKMBLOCK}& 14\\
\end{tabular}\hspace{.3cm}
\begin{tabular}{c c c c}
  Parameter& Name & LH block & LH counter\\
  \hline\hline
  $s_{12}^{\rm(PMNS)}$ & {\tt PMNSs12} & {\tt PMNSBLOCK}& 1\\
  $s_{23}^{\rm(PMNS)}$ & {\tt PMNSs23} & {\tt PMNSBLOCK}& 2\\
  $s_{13}^{\rm(PMNS)}$ & {\tt PMNSs13} & {\tt PMNSBLOCK}& 3\\
  $\delta_{\rm PMNS}$  & {\tt PMNSdel} & {\tt PMNSBLOCK}& 4\\
  \hline
  $s_{12}^{\rm(PMNS')}$ & {\tt PMNSps12} & {\tt PMNSBLOCK}& 11\\
  $s_{23}^{\rm(PMNS')}$ & {\tt PMNSps23} & {\tt PMNSBLOCK}& 12\\
  $s_{13}^{\rm(PMNS')}$ & {\tt PMNSps13} & {\tt PMNSBLOCK}& 13\\
  $\delta_{\rm PMNS'}$  & {\tt PMNSpdel} & {\tt PMNSBLOCK}& 14\\
\end{tabular}
\caption{New physics external parameters of our ALRSM implementation, 
  together with their name and the Les Houches (LH) block and counter
  information allowing to change its numerical value on run time. We recall that
  for consistency, $\kappa < 0$ and the conditions of eqs.~\eqref{eq:copos} and
  \eqref{eq:cstr2} must be satisfied. Those parameters supplement the usual set
  of electroweak inputs given in the LEP scheme, as well as all SM fermion
  masses.}
\label{tab:parameters}
\end{table}

We collect the properties of the new physics fields and external parameters
associated with our \fr\ implementation of the ALRSM model in
tables~\ref{tab:fields} and \ref{tab:parameters}, where we additionally
include properties useful for the user when running any programme relying on our
implementation.

As can be noticed from the tables, the left-handed and right-handed scotinos are
combined to form a Dirac fermion $n_i$ (with $i=1,2,3$ being a generation index)
and the left-handed and right-handed exotic quarks are combined to form a Dirac
fermion $d'_i$ (with $i=1,2,3$ being again a generation index). Whilst all
fermion masses are free parameters of the model (see also
appendix~\ref{app:ferms}), all boson masses are internal (\ie\ are derived
parameters), with the exception of the SM Higgs boson mass $M_{H_0^0}$ (see
appendix~\ref{app:higgs}) and the $Z$-boson mass. As for the SM implementation
included with \fr, our model defines the electroweak sector following the LEP
scheme that is known to yield the minimal parametric uncertainty in the
predictions. The three electroweak inputs are thus the Fermi coupling $\gF$, the
fine structure constant $\alpha$ and the $Z$-boson mass $M_Z$. The gauge and
scalar sectors are then fully defined by fixing nine parameters, that we choose
to be $v'$, $t_\beta$, $\gR$, $\lambda_2$, $\lambda_3$, $\alpha_1$, $\alpha_2$,
$\alpha_3$ and $\kappa$. We recall that the user must ensure that the conditions
of eqs.~\eqref{eq:copos} and \eqref{eq:cstr2} are satisfied when providing the
numerical values of these parameters, and that $\kappa<0$ to avoid tachyonic
charged Higgs bosons.

All other parameters of the gauge and Higgs sectors are then derived as follows.
The vacuum expectation values $v$, $\vL$, $\vR$ and $k$ are obtained from
$\gF$, $v'$ and $t_\beta$,
\be
  v^2 = \frac{1}{\sqrt{2}\gF} \ ,\qquad
  \vL = v\ \cos\beta\ , \qquad
  k = v\ \sin\beta
  \qquad\text{and}\qquad
  \vR^2 = v'^2-k^2 \ .
\ee
As in the SM the $W$-boson mass is derived from the electroweak
inputs,
\be
  M_W^2 = \frac{M_Z^2}{2}
    \bigg[1 + \sqrt{1 - 2\sqrt{2}\frac{\pi\alpha}{\gF M_Z^2}}\bigg] \ ,
\ee
so that eq.~\eqref{eq:mw_mwp} can be used to derive the $SU(2)_L$ gauge
coupling $\gL$. As $e = \sqrt{4 \pi\alpha}$, one can then derive the hypercharge
coupling $\gY$ and the sine and cosine of the electroweak mixing angle $\tw$
from eq.~\eqref{eq:ewmix}, which further allows us to calculate the $B-L$ coupling
constant $\gBL$, the cosine of the $\phw$ mixing angle and the so far neglected
$Z-Z'$ mixing. It is up to the user to verify that his/her choice of input
parameter yields $\tan(2\zw) \lesssim 10^{-3}$. Furthermore, the $W'$- and $Z'$-boson masses are obtained from eqs.~\eqref{eq:mw_mwp} and \eqref{eq:mz_mzp},
and the other parameters of the Higgs potential (\ie, $\mu_1$,
$\mu_2$, $\lambda_1$ and $\lambda_4$) are obtained from eq.~\eqref{eq:minpot}
and eq.~\eqref{eq:lam1}.

In the fermion sector, the various CKM and PNMS matrices are obtained from their
standard expressions in terms of three mixing angles and a phase,
\be\renewcommand{\arraystretch}{1.}
  V = \bpm
   c_{12} c_{13} & s_{12}c_{13}& s_{13}e^{-i\delta}\\
   -s_{12}c_{23}-c_{12}s_{23}s_{13}e^{i\delta} &
    c_{12}c_{23}-s_{12}s_{23}s_{13}e^{i\delta} & s_{23} c_{13}\\
    s_{12}s_{23}-c_{12}c_{23}s_{13}e^{i\delta} &
   -c_{12}s_{23}-s_{12}c_{23}s_{13}e^{i\delta} & c_{23} c_{13}
  \epm \ ,
\ee
where $s_{ij}\equiv \sin \theta_{ij}$ and $c_{ij} \equiv \cos\theta_{ij}$ denote
the sine and cosine of the various mixing angles. Concerning the SM CKM matrix,
we have however traded the input parameters by the usual Wolfenstein parameters
$A$, $\lambda$, $\bar\rho$ and $\bar\eta$,
\be
  s_{12}^{\rm (CKM)} = \lambda \ , \qquad
  s_{23}^{\rm (CKM)} = A\lambda^2 \qquad\text{and}\qquad
  s_{13}^{\rm (CKM)} e^{i\delta_{\rm CKM}} =
    \frac{A \lambda^3 \sqrt{1-A^2\lambda^4} (\bar\rho+i\bar\eta)}
      {\sqrt{1-\lambda^2} \big[1-A^2\lambda^4(\bar\rho+i\bar\eta)\big]} \ .
\ee

\bibliography{ALRM.bib}

\end{document}